%% file: main.tex
\documentclass[sigconf, 9pt]{acmart}

\setcopyright{none}
\renewcommand\footnotetextcopyrightpermission[1]{}
\usepackage{hyperref}

\usepackage[utf8]{inputenc}
\usepackage[color=white]{todonotes}

\usepackage{tabularx}
\usepackage{hhline}
\usepackage{tabu}
\usepackage[font={small,bf}]{caption}
\usepackage{subcaption}
\usepackage{graphicx}
\usepackage{amsmath}
\usepackage[labelformat=simple]{subcaption}
\usepackage{booktabs}
\usepackage{gensymb}

\usepackage{multirow}
\usepackage{makecell}
\usepackage{xspace}
\usepackage{url}
\usepackage{enumitem}
\usepackage{subfiles}
\usepackage{color,soul}

\usepackage{balance}

\colorlet{RED}{red}
\definecolor{pink}{RGB}{219, 48, 122}

\newcommand{\CF}{CarFi\xspace}

\begin{document}

\title{CarFi: Rider Localization Using Wi-Fi CSI}






\author{Sirajum Munir}
\authornote{Equal contribution.}
\authornote{Corresponding author.}
\orcid{0000-0003-0624-2580}
\affiliation{%
    \institution{Bosch Research and Technology Center}
    \city{Pittsburgh}
    \state{PA}
    \country{USA}
    }
\email{sirajum.munir@us.bosch.com}

\author{Hongkai Chen}
\authornotemark[1]
\author{Shan Lin}
\affiliation{%
    \institution{Stony Brook University}
    \city{Stony Brook}
    \state{NY}
    \country{USA}
    }
\email{{hongkai.chen, Shan.X.Lin}@stonybrook.edu}

\author{Shiwei Fang}
\authornotemark[1]
\author{Mahathir Monjur}
\author{Shahriar Nirjon}
\affiliation{%
    \institution{UNC Chapel Hill}
    \city{Chapel Hill}
    \state{NC}
    \country{USA}
    }
\email{{shiwei, mahathir, nirjon}@cs.unc.edu}

\thispagestyle{plain}
\pagestyle{plain}

\begin{abstract}
With the rise of hailing services, people are increasingly relying on shared mobility (e.g., Uber, Lyft) drivers to pick up for transportation. However, such drivers and riders have difficulties finding each other in urban areas as GPS signals get blocked by skyscrapers, in crowded environments (e.g., in stadiums, airports, and bars), at night, and in bad weather. It wastes their time, creates a bad user experience, and causes more CO$_2$ emissions due to idle driving. In this work, we explore the potential of Wi-Fi to help drivers to determine the street side of the riders. Our proposed system is called \CF that uses Wi-Fi CSI from two antennas placed inside a moving vehicle and a data-driven technique to determine the street side of the rider. By collecting real-world data in realistic and challenging settings by blocking the signal with other people and other parked cars, we see that \CF is 95.44\% accurate in rider-side determination in both line of sight (LoS) and non-line of sight (nLoS) conditions, and can be run on an embedded GPU in real-time.
    
\end{abstract}

\maketitle
\pagestyle{empty}


\input{tex/1.introduction.tex}




\input{tex/4.system.tex}

\input{tex/5.algorithm.tex}
\input{tex/6.data_collection.tex}

\input{tex/7.evaluation.tex}

\input{tex/8.discussion.tex}

\input{tex/9.related.tex}

\section{Conclusion}
In this paper, we investigate the feasibility of using Wi-Fi based street side determination of riders from a car to assist drivers to locate their riders. This method can potentially enable a smoother pick-up experience.  Our approach uses a two-antenna Wi-Fi chipset for this purpose. After extracting CSI values, it computes relevant features by leveraging the motion of the car and utilizes a data-driven technique to determine the rider side on an embedded GPU in real-time. By performing extensive evaluation in the real-world in both LoS and nLoS conditions, we see \CF achieves 95.44\% accuracy for estimating the rider side. The approach can potentially be very useful when self-driving cars and robotaxis hit the road.


\balance
{
\bibliographystyle{plain}
\bibliography{carfi} 
}

\end{document}

%% file: tex/1.introduction.tex
\section{Introduction}

As people rely on ride-hailing services, e.g., Uber and Lyft, it becomes increasingly important for drivers and riders to find each other without a hitch. Currently, drivers and riders use smartphones, which rely on GPS or cellular signals, to locate each other while far apart, and require them to recognize each other while nearby. However, in urban cities and areas like downtown, where there are numerous skyscrapers, GPS signals often do not work. In addition, there are places, e.g., in airports, where the drivers need to come indoors (such as parking garages) to pick up riders where the building structure blocks GPS signals. Also, it is challenging to locate the actual rider among many people in crowded environments like stadiums, airports, theatres, and bars. Moreover, the situation can worsen due to lack of visibility, e.g., at night and during bad weather (such as rain, storm, and snow). This issue wastes the time of the riders and drivers, causes more CO$_2$ emissions due to idle driving, causes frustration, and creates a bad user experience.

A recent Uber study shows that every user agreed that they do not like to negotiate the pickup point, and 11 out of 16 users find it hard to give directions to the driver when the user is at a new place \cite{uber_study}.
Based on our discussion with a few drivers and riders, we find that determining the street side of the rider is very crucial.
This is because, if the car is on the other side of the street, the rider sometimes must cross the street, which can be unsafe. Also, the drivers do not want to make a U-turn and realize that they were on the right side in the first place, which leads to a double U-turn. So, we focus on determining the street side of the riders. 


Several solutions have been proposed to improve the rider pick-up experience. For example, the vehicle can use a camera and facial recognition~\cite{raji2021face} to identify the rider and subsequently compute the location. However, facial recognition requires the rider to upload his or her photo, which can be privacy-invasive. Moreover, for facial recognition to work, the rider needs to be within the camera's field of view and occupy enough pixels to be successfully recognized and have good lighting conditions. One can also ask the user to scan the surroundings with his or her phone, and then a server can perform 3D reconstruction~\cite{jin20203d} and matching~\cite{liu2017efficient} to the previously established real-world model to compute the exact location of the rider. However, this is a computation-intensive approach, and this method also requires the world to be digitized and constructed to allow such matching. As commercial products, Uber and Lyft have multicolored LED-based lights for riders to recognize their cars. However, such a solution does not work in broad daylight, and it is a rider-oriented solution, i.e., the rider has to find the car, and the driver does not have much information about the location/side of the rider. Our proposed solution overcomes these limitations.

In this paper, we perform an exploratory study to understand the feasibility of using Wi-Fi to address this problem. We propose to utilize smartphones -- which the riders will mostly like to possess for the ride-hailing booking -- and Wi-Fi-enabled dashcams -- increasingly crucial for safety and legal purposes -- to determine the street side of the rider. We call our proposed system \CF. \CF neither requires the rider to upload any photos of him or her nor a photo of the surrounding area, which protects the rider's privacy, reduces the computation load, and does not depend on lighting conditions. To this end, \CF uses Wi-Fi communications between the rider's smartphone and the dashcam onboard the vehicle. The usage of the dashcam is for the purpose of standalone devices that can be installed on any vehicle, but the proposed system does not exclude vehicles that have Wi-Fi already installed to localize the rider. 

\CF uses a two-antenna Wi-Fi chipset embedded in the dashcam to receive the Wi-Fi packets sent by the smartphone held by the rider. This system does not require any modification to the vehicle and the smartphone. The Wi-Fi packets can be generated by the ride-hailing app, which can share the phone's MAC address (or, a randomized MAC address) through the cloud/server to the vehicle (or, driver's app). Thus, the vehicle can listen to the packets generated from the target phone. The system on the vehicle extracts the Channel State Information (CSI) data from the Wi-Fi chipset. After some preprocessing, it performs sub-carrier selection. Then, it extracts relevant features (amplitude difference between antennas, multipath profile, power delay profile) for rider-side determination. 
Then, the contextual and motion-related features are encoded into a data-driven model (LSTM) to classify whether the rider is on the right or the left side of the vehicle. Our approach only uses CSI amplitude and does not use CSI phase information. Thus, it avoids effort for phase calibration.

This work has the following contributions:

\begin{itemize}
    \item First, we perform a comprehensive exploratory analysis to understand the potential of using Wi-Fi CSI in an automotive environment for shared mobility applications. Our empirical study involves determining the set of features that can effectively work in an automotive environment in both line of sight (LoS) and non-line of sight (nLoS) conditions when a vehicle is being driven and encoding the features into the design and implementation of a data-driven model (LSTM) for estimating the side of the rider using only two antennas and CSI amplitude.
    Our \CF system does not require privacy-invasive personal information from the rider such as a photo, avoids heavy computation on the server, and works in the dark.
    

    \item Second, we set up an infrastructure to collect Wi-Fi CSI from a moving vehicle with a drone-based system for annotating the ground truth location of the vehicle when each packet is received. We collect a dataset of 85 rides with over 568,000 Wi-Fi packets in a realistic and challenging environment, considering both LoS and nLoS, where other people and other parked vehicles block Wi-Fi signals. To the best of our knowledge, this is the first dataset to investigate how Wi-Fi CSI changes over time when the receiver is placed inside a moving car for shared mobility applications.
        

    \item Third, based on evaluation using data collected from the real-world, our results show that \CF is 95.44\% accurate in classifying the rider side in both LoS and nLoS conditions. We also implement several baseline solutions using phase difference and other features and show the superiority of our solution. We also evaluate the execution time of our approach in both powerful and embedded GPUs and show that our solution can be run on an embedded GPU in real-time.
    
    
\end{itemize}

%% file: tex/4.system.tex
\section{\CF Overview}

\begin{figure}
    \centering
    \includegraphics[width=0.48\textwidth]{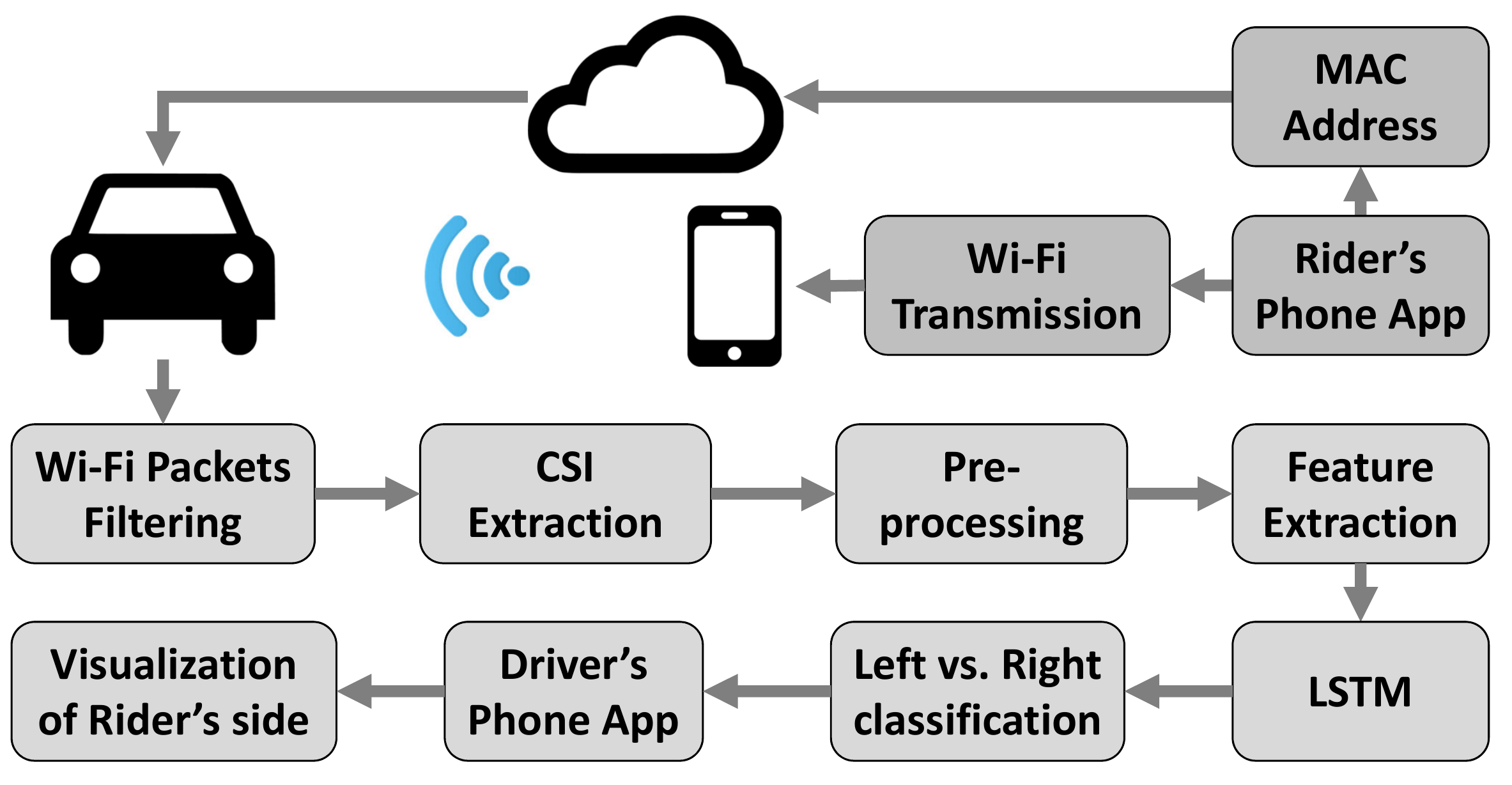}
    \vspace{-0.5em}
    \caption{\CF system overview}
    \label{fig:carfi-sys}
    \vspace{-2em}
\end{figure}

An overview of the \CF system is shown in Figure~\ref{fig:carfi-sys}. When a rider wants to travel to a specific location, s/he uses the ride-hailing phone app on the phone to book a trip. The cloud server of the service providers processes the request and finds a driver. The locations of the vehicle and the rider are determined by their respective location providers, such as GPS on the phone. Once the trip is confirmed, the driver heads toward the rider's location. As the driver arrives within a certain distance, e.g., 0.5 miles from the rider based on the location data, the rider's phone will transmit  Wi-Fi packets at a higher transmission rate as the ride-hailing app controls it. In the meantime, the phone's MAC address is shared with the dashcam via the servers in the cloud. A randomized temporary MAC address can be used to preserve the privacy of the rider. As the vehicle is also within this certain range, the dashcam starts listening for Wi-Fi packets containing the phone's MAC address and filters out other packets. When \CF system receives the Wi-Fi packets with matched MAC address, it extracts the CSI information, performs some pre-processing, and calculates relevant features. Then it feeds the features to an LSTM, which estimates the street side of the rider. Then, this information is passed to the driver's smartphone app from the dashcam for visualization. The data exchange between the phone and the dashcam can be achieved via either Bluetooth or cellular connection (if the dashcam has it).


\section{Challenges}

In this section, we discuss the challenges that \CF system faces for rider side localization in an automotive environment.

\subsection{Automotive Environment}



\begin{figure*}[t]
    \centering
    \begin{subfigure}[b]{0.49\textwidth}
        \centering
        \includegraphics[width=\textwidth]{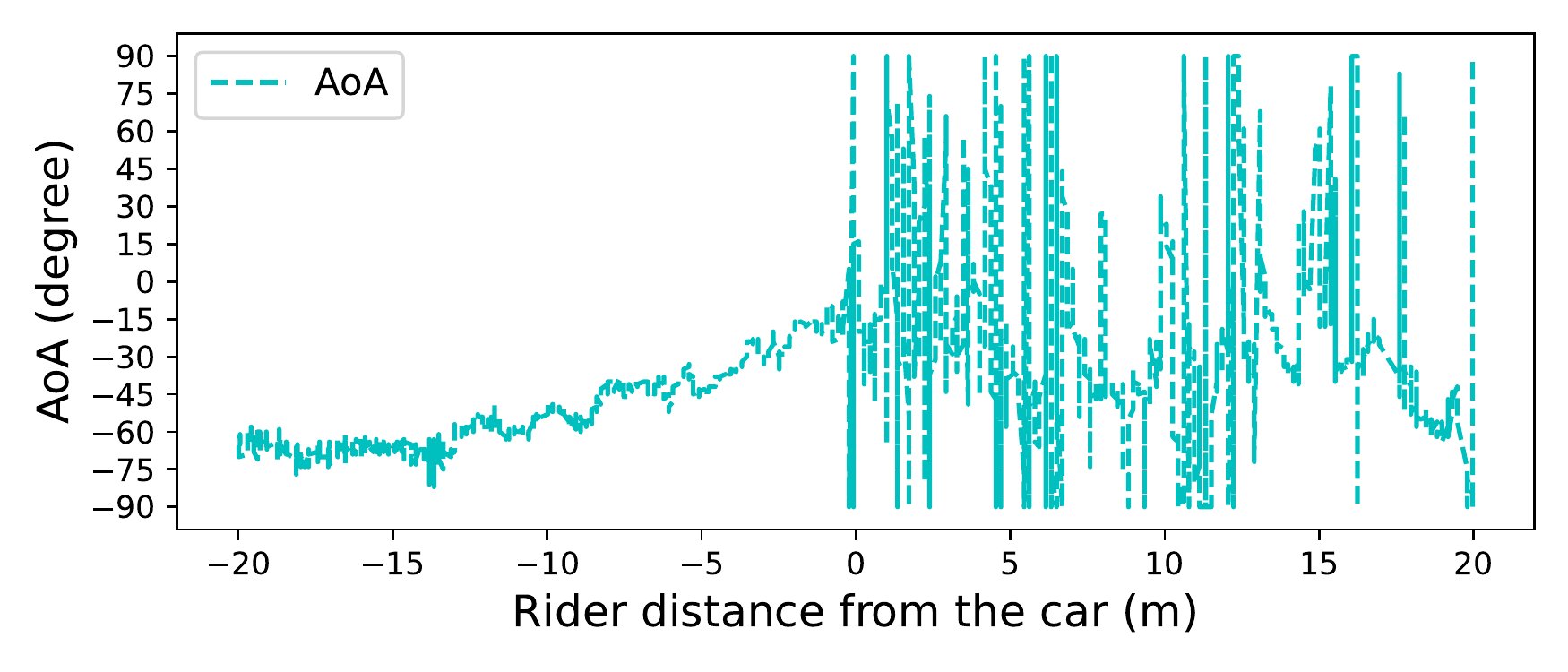}
        \vspace{-2em}
        \caption{AoA estimation of when the rider is in LoS condition.}
        \label{fig:spotfi_1}
    \end{subfigure}
    \hfill
    \begin{subfigure}[b]{0.49\textwidth}
        \centering
        \includegraphics[width=\textwidth]{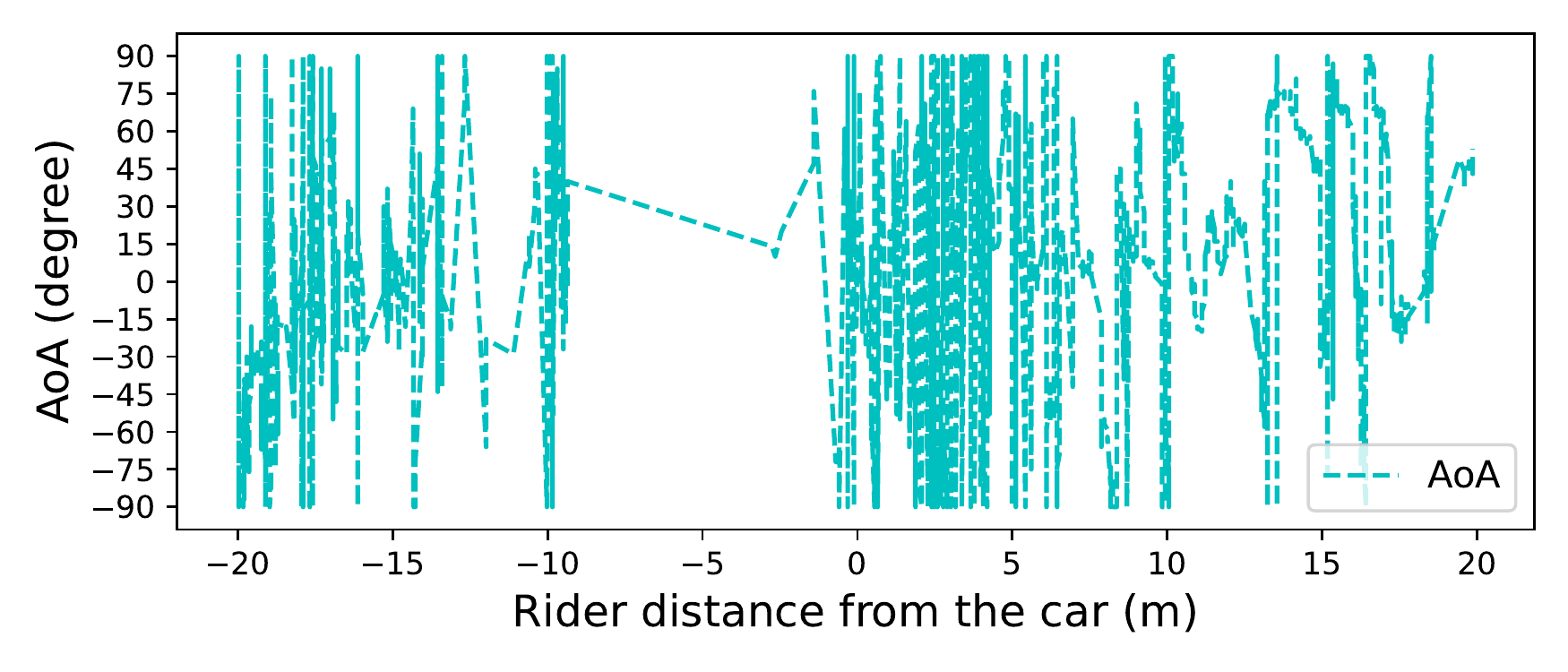}
        \vspace{-2em}
        \caption{AoA estimation of when the rider is in nLoS condition.}
        \label{fig:spotfi_2}
    \end{subfigure}
    \caption{SpotFi AoA in LoS and in nLoS conditions. The rider is in the right side of the car. The expected AoA is 0 to -90\degree.}
    \label{fig:my_label}
    \vspace{-1.5em}
\end{figure*}



When moving Wi-Fi devices from indoor locations to automotive environments, the characteristics of the environment and its effects on the signals change dramatically. One of the biggest issues in an automotive environment is the metal structure of the vehicle body, which can be similar to a Faraday cage. Although the signal of normal radio frequency communication systems has a higher frequency than what the window can block due to its large size, the vehicle's metal surface can still block and redistribute the signal. Unfortunately, there has not been much work to understand how Wi-Fi CSI looks like inside of a vehicle when the vehicle is being driven.



With such a complex RF environment, the current state-of-the-art method, such as SpotFi~\cite{kotaru2015spotfi}, can not accurately estimate the Angle of Arrival (AoA) of the Wi-Fi signal. An example of such an AoA estimation is shown in Figure~\ref{fig:spotfi_1}. The X-axis represents the distance of the rider from the car. The car is coming from the left side of the X-axis, meets the rider in the center, and then leaves. The three antenna arrays are placed at the center of the dashboard of the car, and the AoA should be 0 to -90\degree (0 to 90\degree) when the rider is at the right (left) side. We consider two cases: the rider is standing without anyone blocking the signal (Figure \ref{fig:spotfi_1}), and two other cars and three other people blocking the signal (Figure \ref{fig:spotfi_2}). The rider was on the right side in both cases. We see that in LoS cases, the AoA is relatively stable as the Wi-Fi signal penetrates through the front windshield, but when the car leaves the rider, there is a lot of fluctuation of the AoA as the backside of the car blocks the signal. We observe that when other people and cars block the rider, the AoA is unreliable even when the rider is in front of the car. Since AoA estimation also requires three antennas and phase calibration, we do not use AoA in our approach. 



\subsection{Speed and Time}

We do not expect that the vehicle will approach the rider at highway speed when they are nearby. Instead, we assume that the vehicle will be traveling at a lower speed to be able to stop quickly. Therefore, we assume 10 to 20 miles per hour vehicle speed, which translates to 4.47 to 8.94 meters per second. 
We also consider the transmission range of the Wi-Fi signal to be around 70 to 120 meters in the outdoor environment. If the rider is 70 meters in front of the car, the driver has about 7.83 to 15.66 seconds to stop the vehicle. Given the human response time is about 1 to 1.5 seconds, we determine that the if \CF system takes 3 seconds, it will provide adequate time for the driver to respond and stop safely. 
Smartphones can transmit several hundreds of Wi-Fi packets in a second. However, there could be a burst of packet loss due to non-line of sight (nLoS). In addition, the more time we take to make a decision, the higher accuracy we can offer. Thus, a small window size with a variable number of received packets poses a difficult challenge for rider side determination.


\subsection{Cost}

In order to make the solution practical, we need to use inexpensive antennas and a lightweight computing platform. A simpler solution might use two directional antennas to classify left vs. right. However, we need directional antennas with 180-degree horizontal beamwidth, which is expensive. For example, \cite{directional_antenna} costs \$225 per antenna. Cheaper ones have a smaller beamwidth. For example, \cite{directional_antenna_cheaper} costs \$35.94 per antenna, but has only 66 degrees horizontal beam patterns. Also, such directional antennas are bulky and could obstruct the field of view of the driver more. Adding more antennas also helps in improving the accuracy but also increases the cost of the Wi-Fi chipset and antenna chain. Moreover, the solution needs to be lightweight to be able to run on an embedded GPU or accelerators. Although, such an accelerator would increase hardware cost, a dashcam with such capability could provide additional benefits to the drivers by offering additional services e.g., detecting accidents, violence/aggression in the car and providing necessary support by performing audio-visual analysis.

%% file: tex/5.algorithm.tex
\section{Approach}

In this section, we describe the \CF approach in details.

\subsection{Pre-processing}
When the receiving unit starts to receive Wi-Fi packets, \CF timestamps each packet and keeps all the packets within a window size of 3 seconds for processing together. Then, it uses a stride length of 0.4 seconds to create the next window. 

\subsection{Feature selection}
In this section, we discuss the set of features that we use for left vs. right classification.

\subsubsection{Amplitude difference}
We use Channel State Information (CSI) from only two antennas for the classification. We assume the distance between them is $d$. In our exploratory analysis, we have $d$ = 5.2 cm. CSI contains how the RF signal propagates through the environment as they are being affected during transmission. The CSI data collected at the receiver side contains those affected and encoded in the complex form with amplitude and phase information. Each CSI data point is also the Channel Frequency Response (CFR):

\begin{equation} \label{equ:channel-freq}
    H(f;t) = \sum_n^N a_n(t)e^{-j2\pi f\tau_n (t)}
\end{equation}

Where $a_i(t)$ is the amplitude attenuation factor, $\tau_i(t)$ is the propagation delay, and $f$ is the carrier frequency~\cite{tse2005fundamentals}~\cite{ma2019wifi}. 

\begin{figure}[t]
    \centering
    \begin{subfigure}{.5\textwidth}
        \centering
        \includegraphics[width=\textwidth]{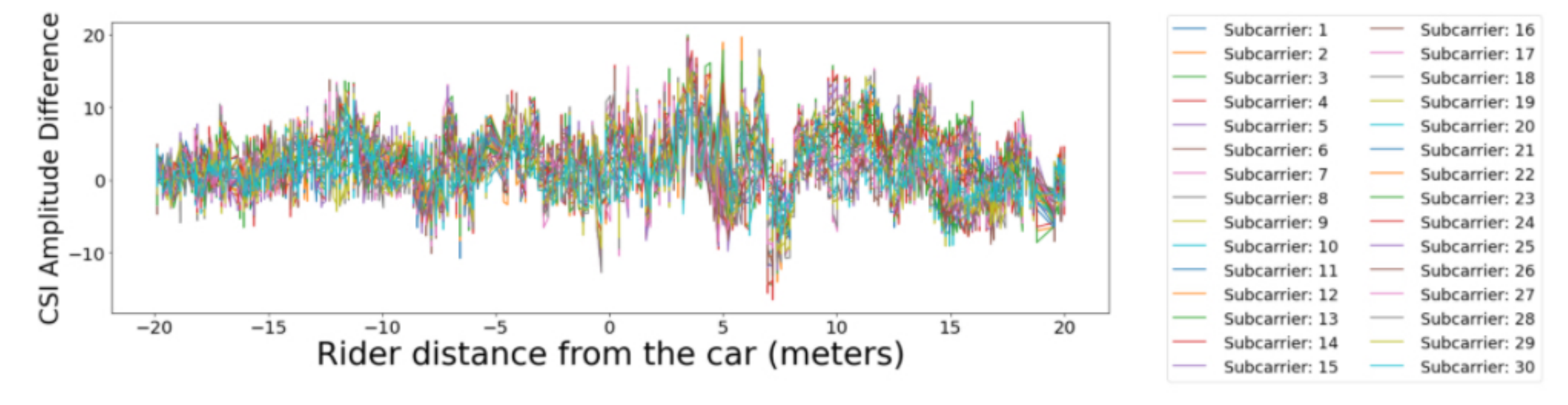}
        \vspace{-2em}
        \caption{Amplitude difference of antennas ($C$ - $A$) when the rider is in LoS.}
        \label{fig:amplitude-difference1}
    \end{subfigure}\quad\quad
    \begin{subfigure}{.5\textwidth}
        \centering
        \includegraphics[width=\textwidth]{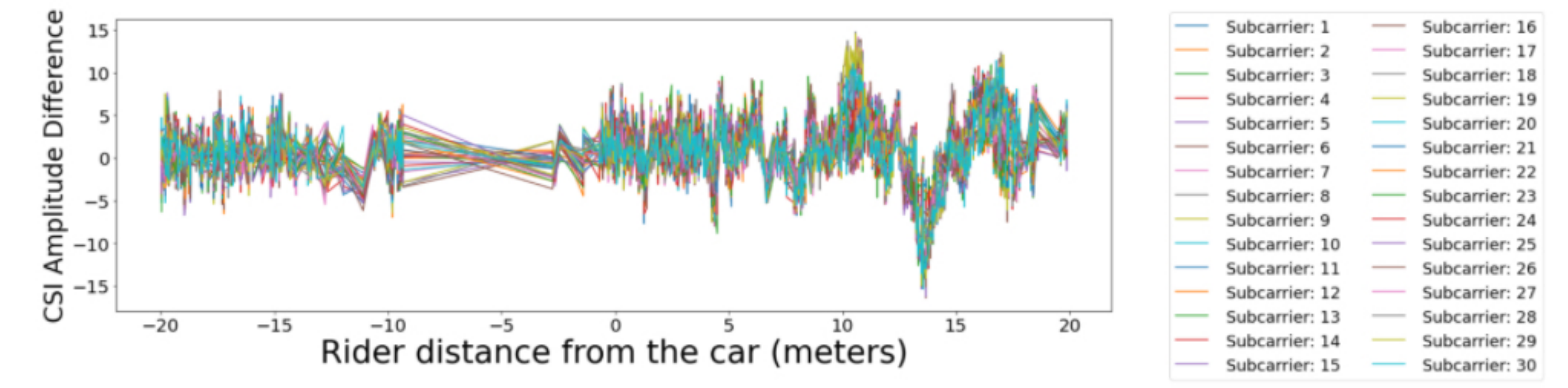}
        \vspace{-1.5em}
        \caption{Amplitude difference of antennas ($C$ - $A$) when the rider is in nLoS.}
        \label{fig:amplitude-difference-nlos}
    \end{subfigure}
    \caption{CSI amplitude difference in LoS and nLoS conditions.}
    \vspace{-1em}
    \label{fig:amplitude-difference}
\end{figure}

Figure \ref{fig:amplitude-difference} shows how CSI amplitude difference between antenna $C$ and antenna $A$ looks like for a portion of a ride for 30 sub-carriers. The rider was on the right side of the car. The X-axis shows the distance of the car with respect to the rider. The car is approaching from the left side of the X-axis, meets the rider at the middle of the X-axis, and then passes the rider after that. When we plot amplitude difference, we plot the CSI amplitude of the antenna $C$ - antenna $A$, where antenna $A$, $B$, and $C$ are placed from left to right parallel to the dashboard (Figure \ref{fig:carfi-ant}). So, a positive value is a good indicator that the rider is on the right side. We see that the amplitude difference values fluctuate over time, and they also vary for different sub-carriers. While  Figure\ref{fig:amplitude-difference1} shows a LoS condition, Figure \ref{fig:amplitude-difference-nlos} shows a nLoS condition where the three other people and two other cars were placed between the rider and the Wi-Fi receiver. We see a burst of packet loss there. As the CSI amplitude varies by subcarriers, instead of relying on all the sub-carriers, we determine the relevant sub-carriers for us that are less prone to noise. 



\subsubsection{Sub-carrier selection} 
\begin{figure}
    \centering
    \includegraphics[width=0.48\textwidth]{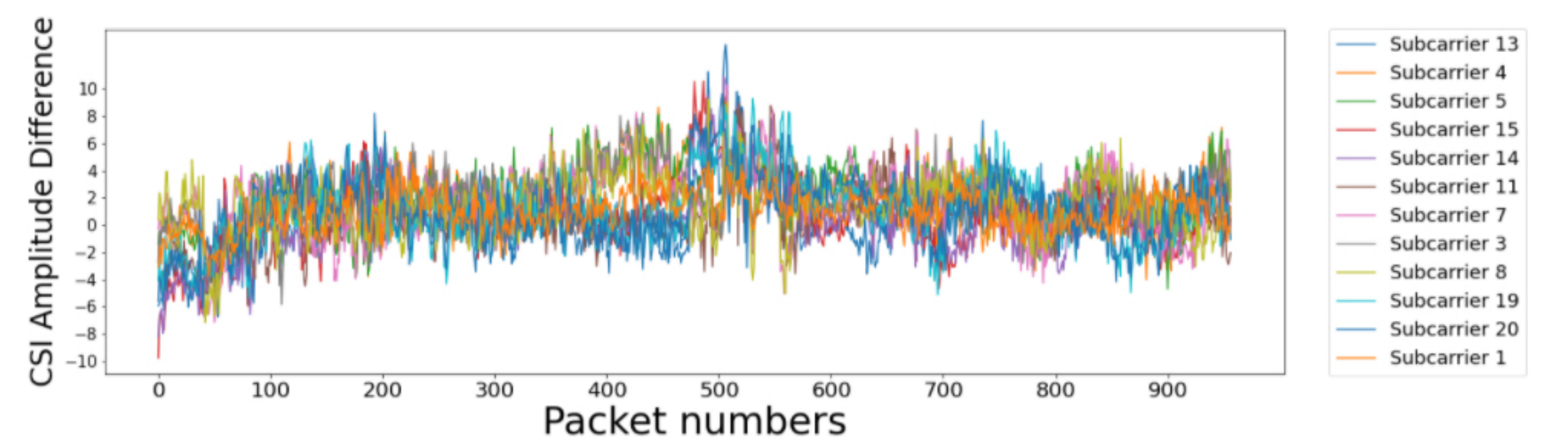}
    \caption{CSI Amplitude difference between antennas  ($C$ - $A$) of 12 selected sub-carriers using Variance-based Subcarrier Selection.}
    \label{fig:amplitude-difference-selected}
    \vspace{-1em}
\end{figure}

Instead of relying on all the sub-carriers, we select sub-carriers that are more resilient to noise. First, we compute the covariance of CSI amplitude of all the subcarriers of antenna C. High covariance between these subcarriers shows they receive effective signal and not the noise. For each subcarrier in antenna C, we select the corresponding subcarrier of antenna A. These subcarriers have similar path properties (e.g., multipath effect, attenuation) and receive correlated CSI data. We vary the number of selected subcarriers from 1 to 30 and choose the number of subcarriers that provide the highest accuracy. Please note that sub-carriers are selected per window of packets. So, different windows may have different sets of sub-carriers. We call this approach Variance-based Sub-carrier Selection (VbSS). When choosing $N$ subcarriers, we choose $N-1$ sub-carriers using VbSS and add the first subcarrier. Figure \ref{fig:amplitude-difference-selected} shows the amplitude difference of the 12 selected subcarriers of a window when the rider is on the right side of the car.

\subsection{Multipath Profile}

\begin{figure}[t]
    \centering
    \begin{subfigure}{.5\textwidth}
        \centering
        \includegraphics[width=\textwidth]{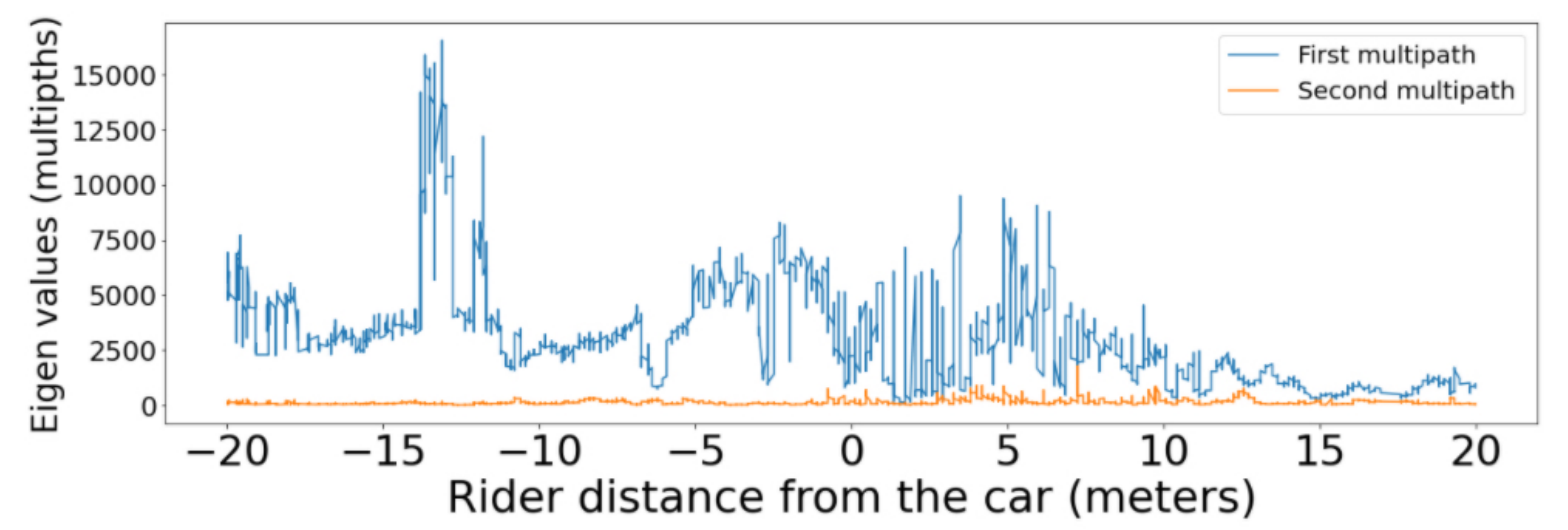}
        \vspace{-1.7em}
        \caption{First and second multipaths of when the rider is LoS condition.}
        \label{fig:multipath-profile1}
    \end{subfigure}\quad\quad
    \begin{subfigure}{.5\textwidth}
        \centering
        \includegraphics[width=\textwidth]{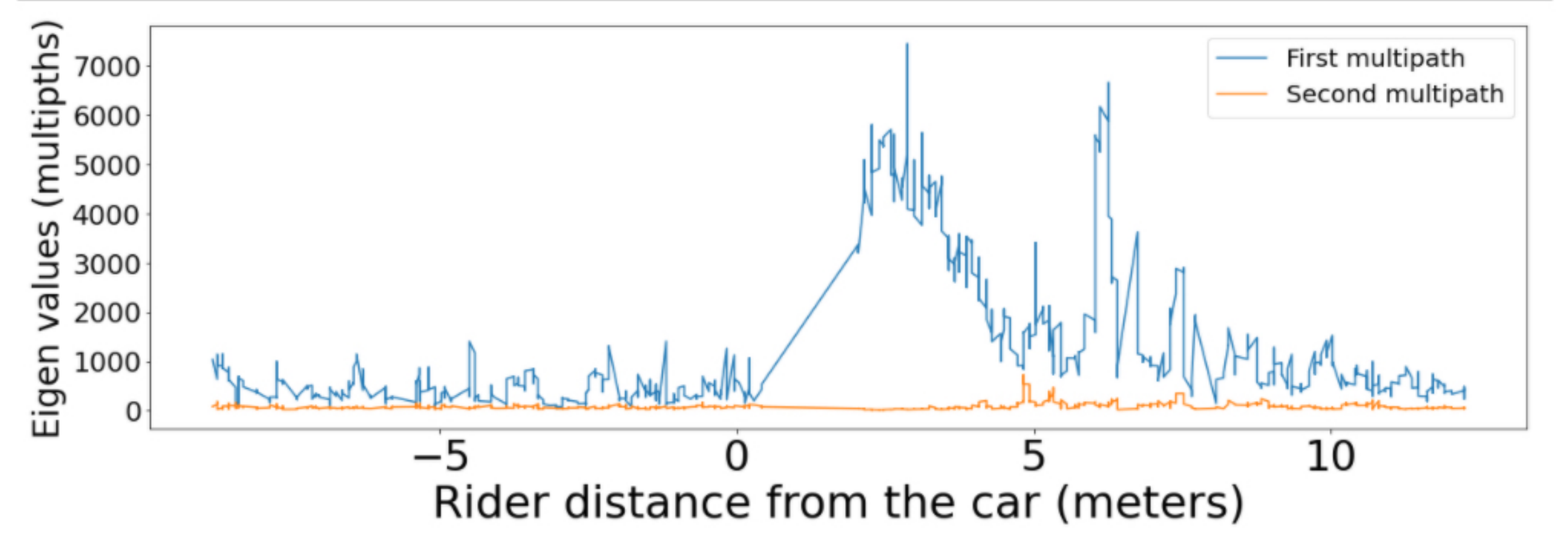}
        \vspace{-1.7em}
        \caption{First and second multipaths of when the rider is nLoS condition. Three other people and two other cars are blocking the signal.}
        \label{fig:multipath-profile2}
    \end{subfigure}
    \caption{Multipath profile in LoS and nLoS conditions.}
    \vspace{-1em}
    \label{fig:carfi-multipath-profile}
\end{figure}

Since Wi-Fi CSI data contains multipath attenuation caused by the environment, the multipath profile extracted from the CSI data can be very useful in location estimation. It can effectively provide whether the rider is in LoS or nLoS conditions. To extract the multipath profile of the CSI data, we explore how MUSIC~\cite{schmidt1986multiple} and SpotFi~\cite{kotaru2015spotfi} algorithms extract signals and estimate their Angle of Arrival. Inspired by this, we first isolate multiple possible signals by performing Eigen decomposition of matrix $XX^H$, where $X$ is the CSI measurement, and $X^H$ is the conjugate transpose of $X$. The eigenvectors and eigenvalues can be used as features as they are affected by the environment and the vehicle. We take the top two dominant multipaths and plot them in  Figure~\ref{fig:carfi-multipath-profile} with both LoS and nLoS conditions, and the rider was on the right side. The X-axis in both figures shows the distance from the car as the car is approaching the rider from the left side of the axis. Figure \ref{fig:multipath-profile1} represents the case when the rider is in a LoS condition. We see that as the car approaches, there is a significant difference between the first and the second multipath. However, as the car leaves the rider, the backside of the car blocks the signal and causes nLoS conditions, and hence the difference between the top two multipaths decreases significantly. Figure \ref{fig:multipath-profile2} represents the case when the rider is in a nLoS condition. There were three other people and two other cars blocking the signal. As a result, the first and second dominant multipath is closer from the beginning. However, when the car passes the rider, it gets a line of sight for a brief moment, and hence the difference between the first and the second multipath becomes more prominent. We take the top two Eigenvalues representing the top two dominant multipaths computed using CSI values of two antennas as features for classification.

\subsection{Power Delay Profile}

Power Delay Profile (PDP) describes the power level associated with each multipath along with the propagation delays. However, due to the limited bandwidth of the Wi-Fi channels, the path length resolution is not very precise. For our 802.11ac 40 MHz channel, the path length resolution is 7.5m. But it can be helpful for coarse-grained mobility tracking over time and provide contextual information regarding LoS and nLoS.

\begin{figure}[t]
    \centering
    \begin{subfigure}{.48\textwidth}
        \centering
        \includegraphics[width=\textwidth]{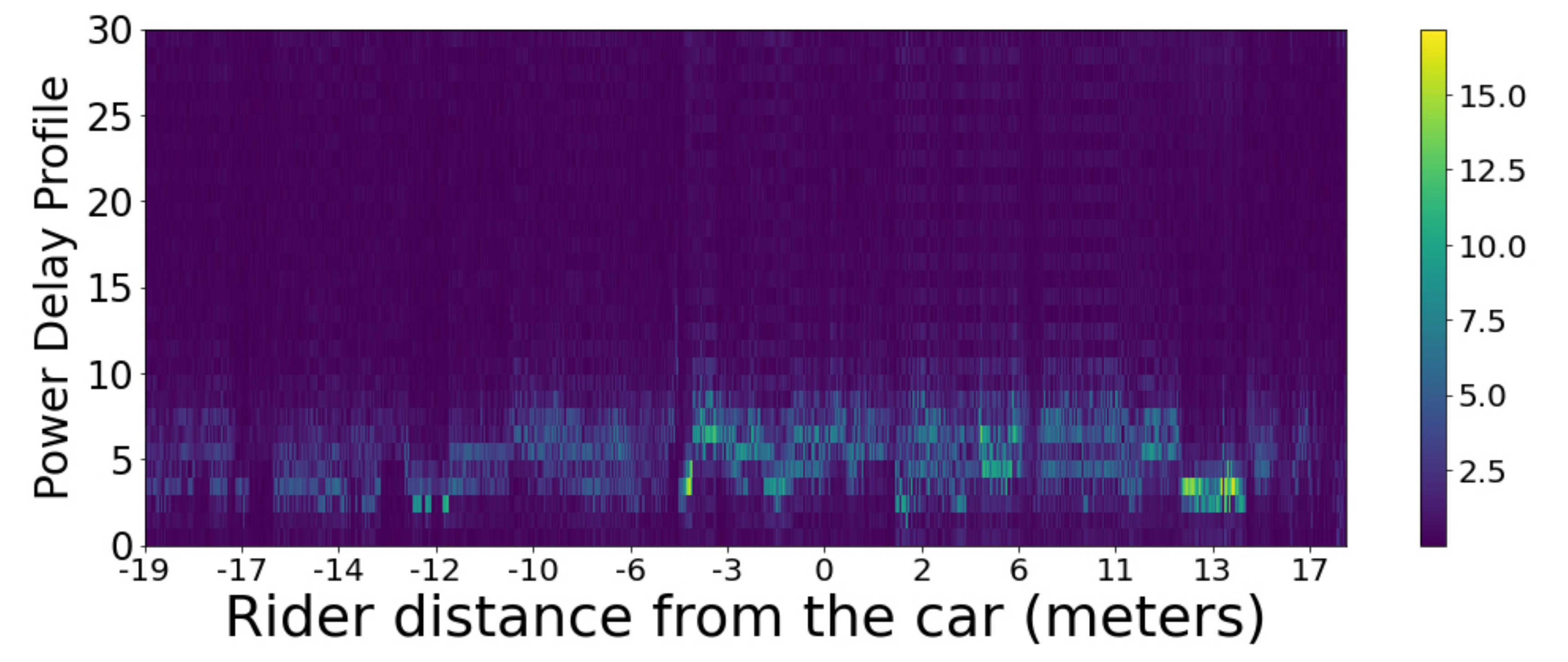}
        \caption{Power delay profile of when the rider is LoS condition..}
        \label{fig:pdp1}
    \end{subfigure}\quad\quad
    \begin{subfigure}{.48\textwidth}
        \centering
        \includegraphics[width=\textwidth]{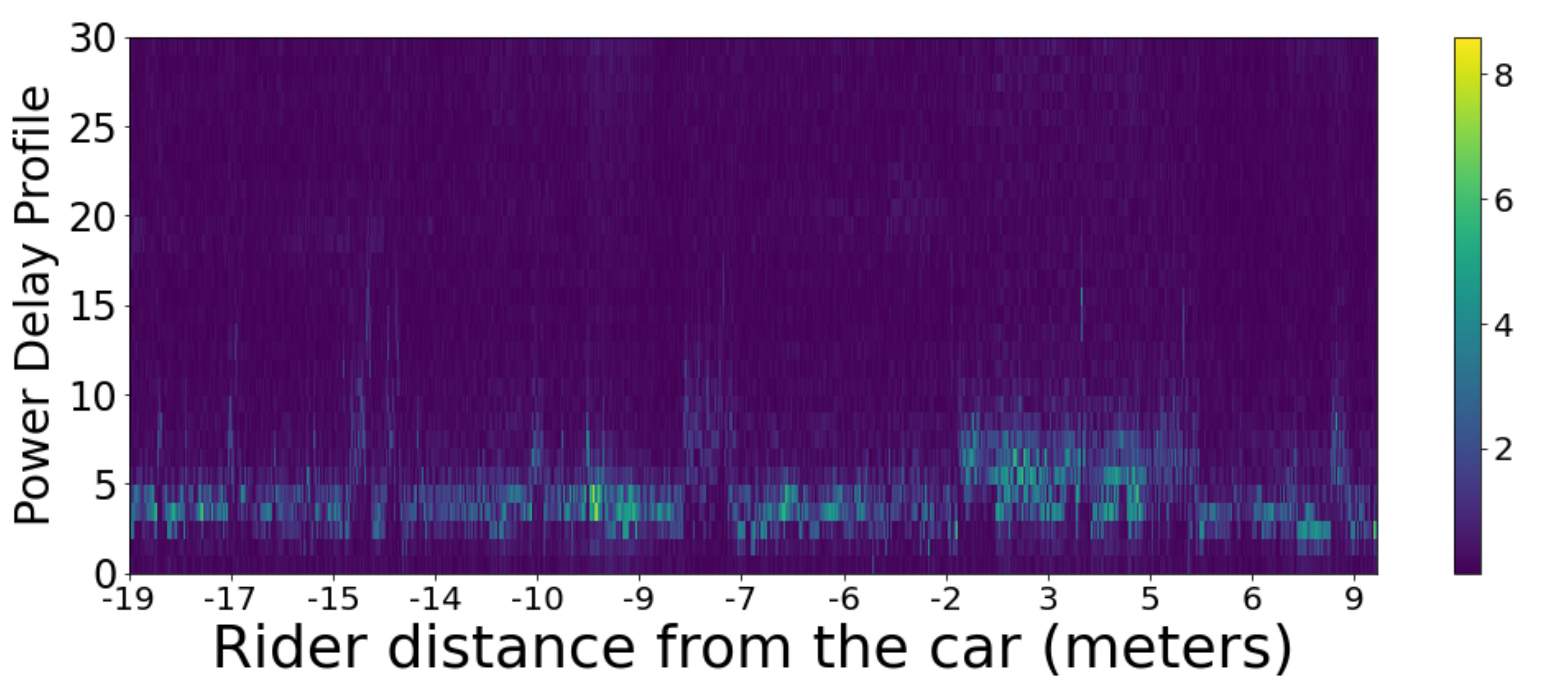}
        \caption{Power delay profile of when the rider is nLoS condition.}
        \label{fig:pdp2}
    \end{subfigure}
    \caption{Power delay profile in LoS and nLoS conditions.}
    \vspace{-1em}
    \label{fig:pdp}
\end{figure}

When the Wi-Fi chipset measures the channel frequency response as written in Equation~\ref{equ:channel-freq}, instead of measuring continuously, it samples the response at discrete frequency points \(f = f_0 + k\Delta f\), where k is the sub-carrier index and \(\Delta f = 312.5kHz\)~\cite{xie2018precise}. Since Equation~\ref{equ:channel-freq} is in the frequency domain, by applying Inverse Fourier Transform, we can get the response in the time domain which is also the Channel Impulse Response (CIR):

\begin{equation}
    f(t) = \sum_{n}^N a_n \delta(t-\tau_t)
\end{equation}

where \(a_n\) and \(N\) is the same as in Equation~\ref{equ:channel-freq} and $\delta(\cdot)$ is the delta function. By calculating the norm $\|f(t)\|_2$ of the Channel Impulse Response $f(t)$, we can get the Power Delay Profile. Each of the signal samples in the Channel Impulse Response correlates to different multipath as their time to travel from the transmitter to the receiver differs due to differences in the traveled length. By considering the IFFT theory, the time resolution \(\Delta \tau\) is related to the sampling resolution \(\Delta f\) mentioned above. While increasing the number of bins in IFFT, the actual resolution does not change. As such, we set our IFFT bins to the number of subcarriers, which is also the frequency sampling resolution. For our collected data, 30 subcarriers are reported for each antenna. By using two antennas, we obtain 60 PDP values as features per Wi-Fi packet. We show the PDP values from one antenna in LoS condition in Figure \ref{fig:pdp1}, and in nLoS condition in Figure \ref{fig:pdp2}, where there are three people and two cars blocking the signal between the rider and his car. In both cases, the rider was on the right side. We see how the PDP values are changing as the car approaches the rider from the left side of the X-axis and passes him. The PDP values do not necessarily tell if the rider is on the left or right side but help contextualize the packets of similar distance, and in LoS/nLoS conditions to provide additional information to the classification model.

\section{Classification}

We consider different classifiers to classify the side of the rider (left vs. right), including k-Nearest Neighbor (kNN), Decision Tree (DT), and Support Vector Machine (SVM). In addition, we design a Long Short-Term Memory (LSTM) neural network classifier by effectively integrating all the features. Now, we describe the design of an LSTM and how we encode the relevant contextual and motion-related features. 


As the vehicle approaches the rider, the motion of the vehicle, as well as the distance between the transmitter (the phone held by the rider) and the receiver (Wi-Fi receiver on the vehicle), provides additional features in the time domain. For example, Wi-Fi signal differences between different antennas can vary across time. There are also Wi-Fi signal differences on the same antenna with different transmitter and receiver distances. Unlike neural network architectures such as Fully-Connected Neural Network and Convolutional Neural Network (CNN), LSTM can better encode time series data with its feedback connections to remember values over arbitrary time intervals. Thus it can exploit the temporal features introduced by the vehicle's motion. While traditional classifiers like k-NN, DT, and SVM can capture features at a single time step, they lack the ability to take into account the temporal features as the signal is coming from either left or right in both LoS and nLoS situations. 


The general execution of LSTM is described in equations below:

\vspace{-2em}
\begin{align}
    i_t &= \sigma(W_{ii}x_t + b_{ii} + W_{hi}h_{t-1} + b_{hi}) \label{equ:lstm1}\\
    f_t &= \sigma(W_{if}x_t + b_{if} + W_{hf}h_{t-1} + b_{hf}) \label{equ:lstm2}\\
    g_t &= tanh(W_{ig}x_t + b_{ig} + W_{hg}h_{t-1} + b_{hg}) \label{equ:lstm3}\\
    o_t &= \sigma(W_{io}x_t + b_{io} + W_{ho}h(t-1) + b_{ho}) \label{equ:lstm4}\\
    c_t &= f_t \odot c_{t-1} + i_t \odot g_t \label{equ:lstm5}\\
    h_t &= o_t \odot tanh(c_t) \label{equ:lstm6}
\end{align}
\vspace{-1.5em}

The main advantage of LSTM to other neural networks in temporal feature understanding is the memory cell $c_t$, which is used to accumulate state information in each time step. To decide what to remember and forgot, Equation~\ref{equ:lstm1} and ~\ref{equ:lstm2} calculate the input gate and forget gate value, respectively. The input gate $i_t$ decides which information (which is calculated by Equation~\ref{equ:lstm3}) is saved to the memory cell. On the other hand, the forget gate $f_t$ control which part of the previous cell status could be forgotten. With these calculations, we determine what is the new memory cell status through Equation~\ref{equ:lstm5}. Additionally, how the memory cell $c_t$ propagates to the final state or output $h_t$ (through Equation~\ref{equ:lstm6}) is controlled by the output gate $o_t$ (calculated at Equation~\ref{equ:lstm4}). This design allows the LSTM to take into account previous state information and can be self-learned through the training process. In these equations, $x_t$ is the data at time step $t$, $b$ is the bias in each network connection, the upper case $W_i$ and $W_h$ represents the matrices of the weight of the input data and recurrent connection, respectively, and $\odot$ is the Hadamard product.



\begin{figure}
    \centering
    \includegraphics[width=0.43\textwidth]{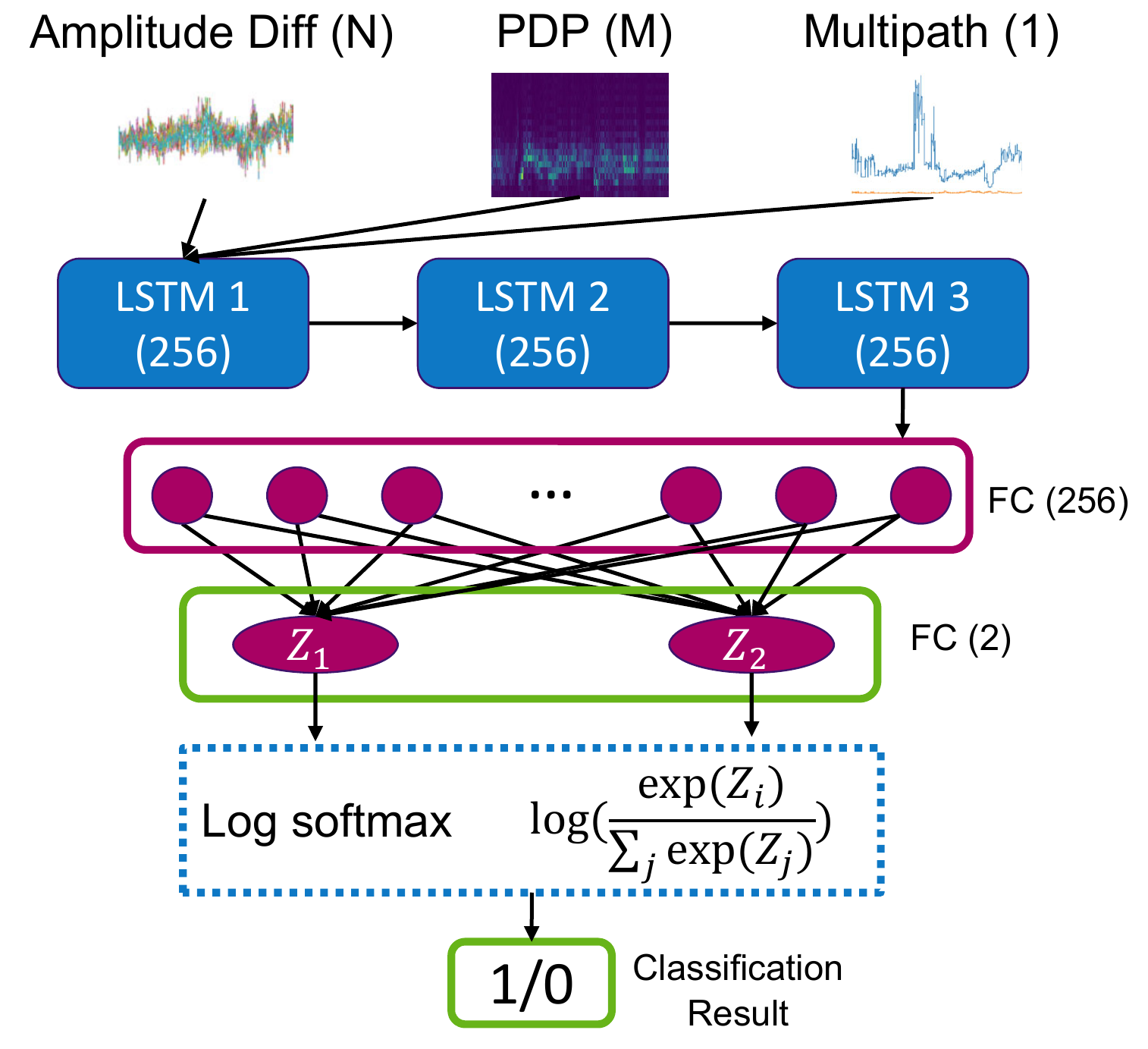}
    \vspace{-0.7em}
    \caption[LSTM]{LSTM architecture}
    \vspace{-2em}
    \label{fig:LSTM-architecture}
\end{figure}

The architecture of LSTM is shown in Figure \ref{fig:LSTM-architecture}. It has an input size of N. 3 LSTM layers are stacked with 256 hidden units. They are followed by a linear layer with an input size of 256 and an output size of 2. A Softmax layer is added after the linear layer. The training uses a cyclic learning rate with a 5e-4 initial learning rate with maximum epochs of 650 with the patience of 200. For loss function, we use cross-entropy loss. We have a dropout of 0.5 in the LSTM layers. 

The sequence length in LSTM for each sample (or, a window) needs to be the same. However, we observe burst of packet losses in nLoS conditions. As a result, the number of packets varies from windows to windows (in 3 seconds). Hence, the length of the LSTM sequence needs to be determined. We take the median of the number of packets of the windows of the training set, which is 855 packets and set that the sequence length of LSTM. If there are more packets, then we ignore the rest. If there are fewer packets, then we perform zero padding at the end of the sequence. In this way, we actually take 1.5 seconds of Wi-Fi packets half of the time for the classification.

Before feeding the CSI amplitude difference, power delay profile, and multipath profile features to LSTM, we normalize them. This is important to make sure different features with different scales (especially the dominant multipath) do not force the network to weigh differently. So, the features from the multipath profile and power delay profile need to be crafted in a way that even after normalization, the distinction of LoS and nLoS does not disappear. In order to ensure that, we create just one feature using the multipath profile by dividing the magnitude of the dominant multipath with that of the less dominant multipath. For the power delay profile, adding 60 input channels to LSTM may cause over-fitting. So, we apply Principal Component Analysis of the 60 PDP features and take the top M principal components to feed to the network. We vary M from 3 to 5 and show the results in the Evaluation section. After this process, the normalization retains the LoS and nLoS distinction and reduces the number of input channels to LSTM to reduce overfitting.

%% file: tex/6.data_collection.tex
\section{Data Collection}

\begin{figure*}[ht]
    \centering
    \includegraphics[width=0.98\textwidth]{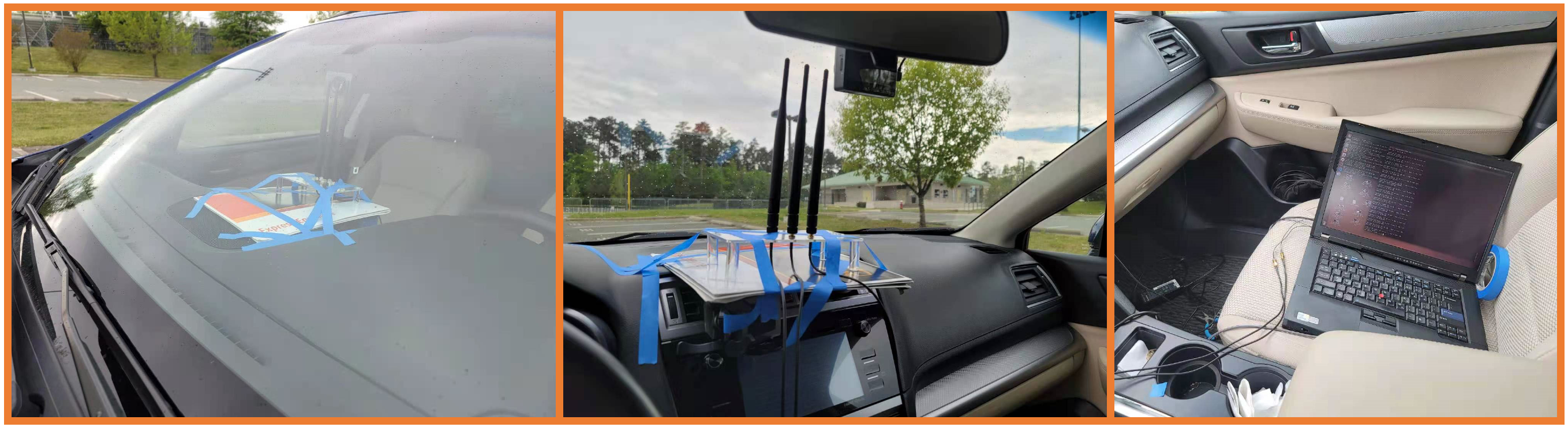}
    \caption{In-vehicle system setup. The left picture shows the antenna viewed from outside. The middle picture shows Wi-Fi antennas placed on the central dashboard. Antennas from  the left to the right are labeled as $A$, $B$, and $C$. The right picture shows a laptop with an Intel 5300 NIC and connected with antennas through cables.}
    \vspace{-1.3em}
    \label{fig:carfi-ant}
\end{figure*}

\begin{figure}[t]
    \centering
    \begin{subfigure}{.23\textwidth}
        \centering
        \includegraphics[width=\textwidth]{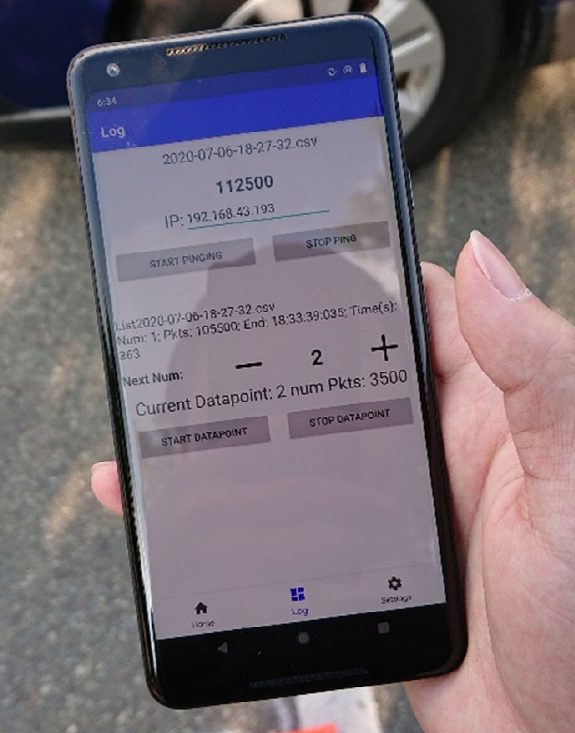}
        \caption{Android App developed to generate Wi-Fi traffic.}
        \label{fig:phone-app}
    \end{subfigure}
    \begin{subfigure}{.23\textwidth}
        \centering
        \includegraphics[width=\textwidth]{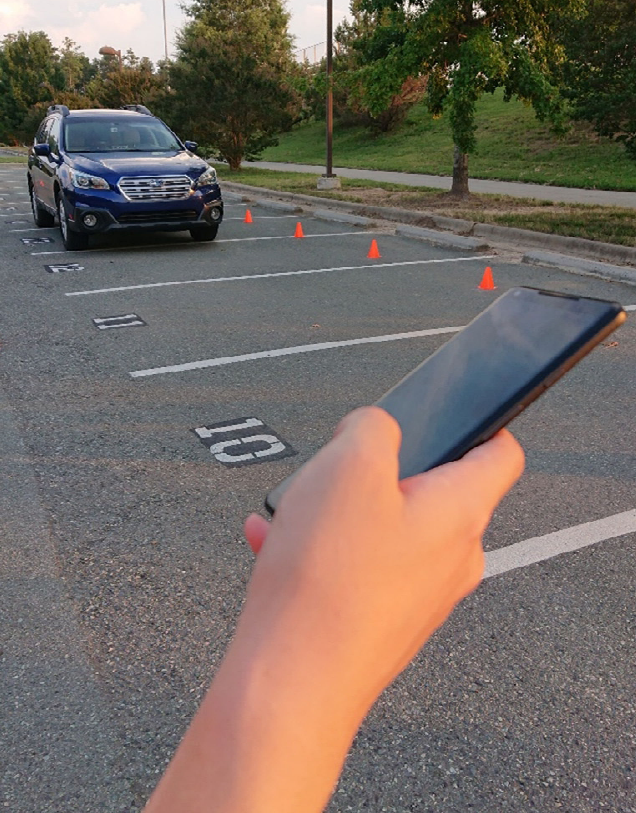}
        \caption{View of how the phone is held relative to the car.}
        \label{fig:phone-car}
    \end{subfigure}
    \caption{Data collection equipment and environment.}
    \vspace{-2em}
    \label{fig:carfi-collet}
\end{figure}

\subsection{System Setup}

To extract the Wi-Fi data in an automotive environment, we utilize a laptop with an Intel 5300 Wi-Fi Network Interface Card (NIC) for portability, as shown in Figure~\ref{fig:carfi-ant}. We use the Linux CSI tool~\cite{halperin2011tool} to collect PHY layer CSI information from received Wi-Fi packets. The car is driven with this set up for receiving Wi-Fi packets about 10-20 miles per hour. The three antennas are placed in the dashboard. We mark them as A (leftmost), B (middle), and C (rightmost), when viewed from inside of the vehicle. Although we collect data with 3 antennas, we use only two antennas for our approach (antennas A and C). On the rider side, the rider stands with a Pixel 2 XL phone that serves as an Access Point (AP) at 5 GHz to which the laptop is connected to. An Android app from the phone generates Wi-Fi traffic by pinging the laptop as shown in Figure~\ref{fig:phone-app}, that we developed and can achieve a packet transmission rate of up to 
350 packets per second.

\subsection{Collected Dataset}

In order to consider realistic scenarios with LoS and nLoS conditions, we collect data of 85 rides under five different conditions: (a) only rider standing, (b) people standing on both sides of the rider, (c) two other people blocking the signal, (d) two other parked cars blocking the signal, and (e) two other cars and three other people blocking the signal. We collect data when the rider is on the left and right sides of the car in all these conditions. Table~\ref{tab:dataset} shows the number of rides under different conditions. Figure \ref{fig:carfi-three-p-car} shows a drone image of the case (e), where the rider is standing, and three people and two cars are blocking the signal. The Wi-Fi receiving unit is in the blue car that was being driven from the left to the right side.

\begin{table}[t]
\centering
\caption{Distribution of 85 rides under different conditions.}
\label{tab:dataset}
\resizebox{\linewidth}{!}{
\begin{tabular}{@{}lcc@{}}
\toprule
                                              & Rider left side & Rider ride side \\ \midrule
Only Rider                                    & 7               & 6               \\ \midrule
People both sides of the rider (no car)       & 5               & 6               \\ \midrule
Two other people blocking signal              & 13              & 14              \\ \midrule
Two cars blocking signal (no other people)    & 10              & 12              \\ \midrule
Two cars and three other people blocking signal & 6               & 6               \\ \bottomrule
\end{tabular}
}
\vspace{-1em}
\end{table}


We split the dataset into training (60\%), validation (20\%), and testing (20\%). We do this per for each condition and for each side of the rider. For example, when the rider is at the left side and two cars blocking the signal, we have 10 such rides. We take CSI data of 6, 2, and 2 rides for training, validation, and testing, respectively. In that way, the test set has data of disjoint rides and under all conditions. For each ride, we split the sequence of CSI values into a 3 seconds window with 0.4 seconds stride length. This gives us 1032 windows for training, 286 windows for validation, and 285 windows for testing.

\subsection{Ground Truth Collection}

\begin{figure}
    \centering
    \includegraphics[width=0.48\textwidth]{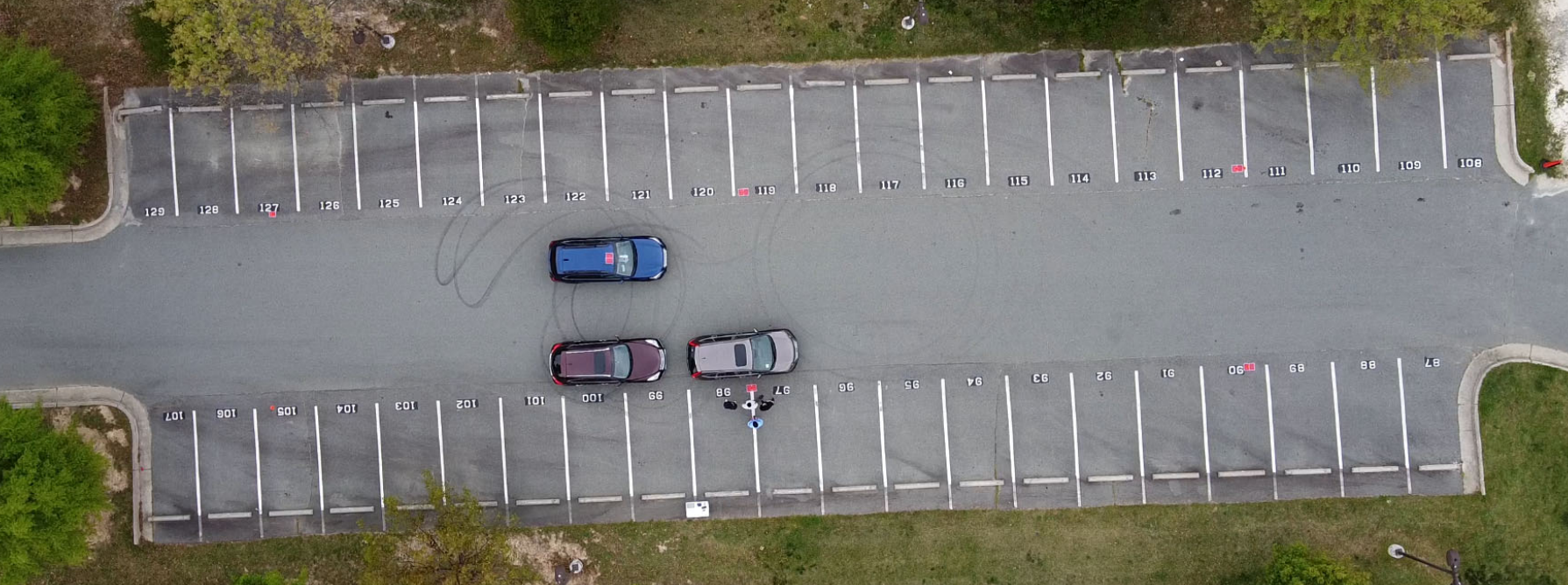}
    \caption{Drone image of three people and two vehicles in between the Wi-Fi transmitter and receiver.}
    \vspace{-1em}
    \label{fig:carfi-three-p-car}
\end{figure}

In order to collect the ground truth of whether the rider is at the left or right side of the car, one can just record the timestamps of received packets for each side of the rider. However, we would like to collect the (x,y) location of the car when each packet was received to have a better understanding of how the CSI changes when the vehicle approaches the rider and leaves the rider at each side. In order to achieve this goal, we use an off-the-shelf consumer drone hovering above the data collection site to record the process. An example frame from the recorded video is shown in Figure~\ref{fig:carfi-three-p-car}. Before the data collection, we first determine landmark locations (e.g., the rider's location) and four positions that can form a rectangle area with tape-measured ground truth coordinates. Next, we place solid red-colored papers at each location and on top of the car to enable simple color-based pixel tracking through color thresholding. In the recorded video, we use the four locations to perform Homography transformation so that the pixel plane and real-world plane are parallel. This transformation creates a straightforward translation from the pixel coordination system to the real-world coordination system through scaling. Then we can track the vehicle in the pixel domain and interpolate the real-world (x,y) location through the translation. Prior to each data collection, we also time-synchronize the Android phone, the laptop with the Intel Wi-Fi chipset, and the drone. The time-synchronization between the phone and drone is achieved by capturing the phone's time with millisecond accuracy at the beginning of each drone's video; thus, we can calculate the timestamp based on the frame rate and a reference frame that has the phone's time clearly recorded. We also capture a screenshot with both the phone's time (through the laptop's camera) and the laptop's time displayed with millisecond accuracy; thus, the time difference between them can be easily calculated. We apply these time offsets to change the timestamp recorded on the laptop and the drone to match the time on the phone for time synchronization.

%% file: tex/7.evaluation.tex
\section{Evaluation}

In this section, we estimate the accuracy of \CF and compare it with state-of-the-art methods. We investigate the effect of antenna spacing, subcarrier selection, and window size on the performance of the solution. Also, we estimate its execution time, and range in both LoS and nLoS conditions.

\subsection{Accuracy}

To the best of our knowledge, there is no state-of-the-art Wi-Fi-based rider side determination technique. So, we implement a few baseline methods to compare with our approach in terms of accuracy.

\textbf{Baseline 1: CSI phase difference based approach:}
Although our approach does not require phase calibration, in order to investigate and compare with a phase difference based approach, we perform phase calibration of the antenna chains of the Intel 5300 chipset attached with the laptop using a method similar to~\cite{xiong2013arraytrack}. We use another laptop with Intel 5300 chipset to transmit Wi-Fi packets through an RF splitter, where all \CF three receiver antennas are connected to the RF splitter's output as shown in Figure~\ref{fig:phase-calibration}. These three antennas should receive the Wi-Fi signal at the same time. However, due to the slight path distance difference within the RF splitter, we switch the receiver antennas' connecting locations and record the phase information in each connection combination to eliminate the difference introduced by the RF splitter. By removing the offset we measured, we correct the antenna phase offset in our collected data. The system also introduces Sampling Time Offset (STO) and Sampling Frequency Offset (SFO) as the sampling clocks and frequencies are unsynchronized between the receiver and the transmitter. 
We follow SignFi~\cite{ma2018signfi}  to remove STO and SFO through multiple linear regression. 

\begin{figure}[t]
    \centering
    \begin{subfigure}{.5\textwidth}
        \centering
        \includegraphics[width=\textwidth]{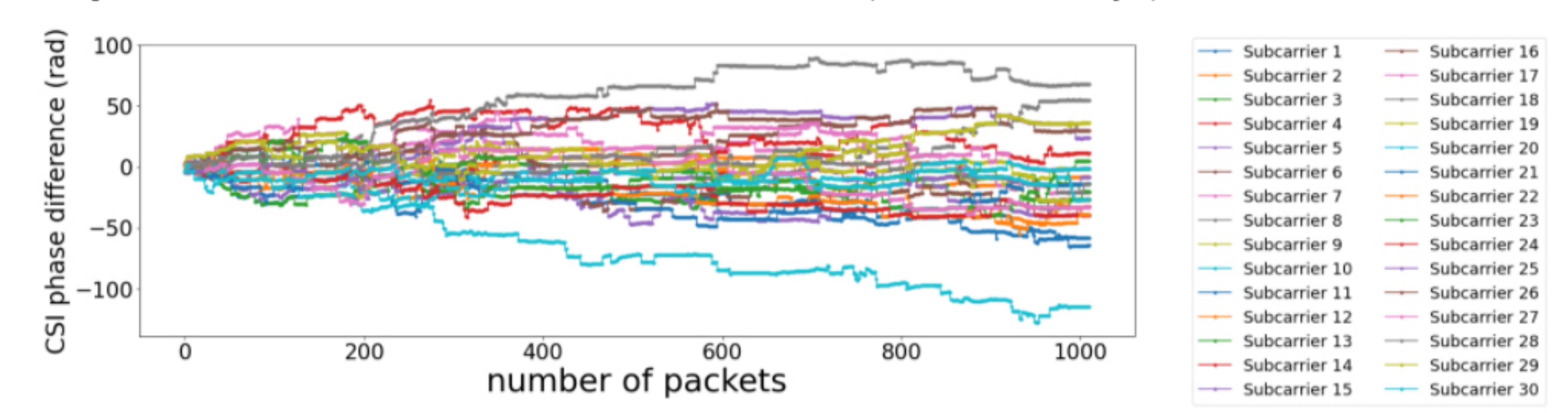}
        \caption{Unwrapped phase difference of antennas ($C$ - $A$) in LoS condition.}
        \label{fig:phase-difference1}
    \end{subfigure}\quad\quad
    \begin{subfigure}{.5\textwidth}
        \centering
        \includegraphics[width=\textwidth]{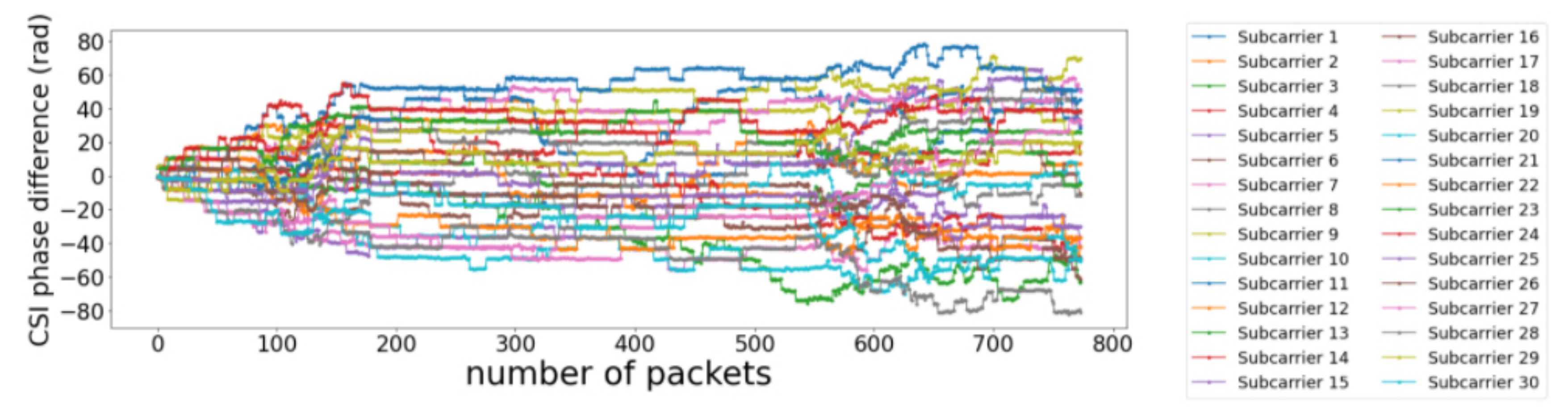}
        \caption{Unwrapped phase difference of antennas ($C$ - $A$) in nLoS condition.}
        \label{fig:phase-difference2}
    \end{subfigure}
    \caption{Unwrapped phase difference in LoS and nLoS conditions when the rider is in the right side.}
    \vspace{-1em}
    \label{fig:phase-difference}
\end{figure}

We use only antennas $A$ and $C$, and estimate the phase difference by subtracting unwrapped phase $A$ from unwrapped phase $C$ of each window. Ideally, the phase difference should be positive (negative) when the rider is on the left (right) side. But for 30 different sub-carriers, the patterns vary significantly. As an example, we show the phase difference when the rider is at the right side in LoS condition and when he was blocked by two cars in  Figure \ref{fig:phase-difference}.

Since the unwrapped phase difference change over time, we consider four different ways to compute the features to capture phase difference between antennas:

\begin{figure}
    \centering
    \includegraphics[width=0.45\textwidth]{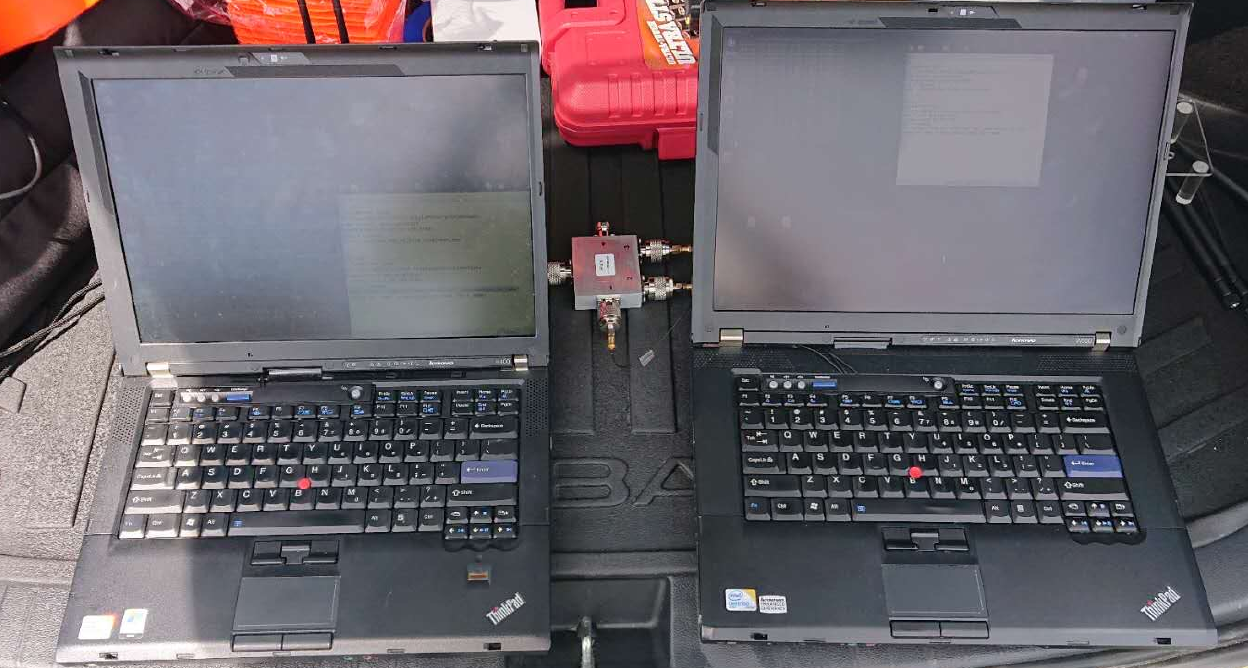}
    \caption{Antenna chain phase calibration of the Intel 5300 chipset.}
    \vspace{-1.5em}
    \label{fig:phase-calibration}
\end{figure}

\begin{figure}[t]
    \centering
    \includegraphics[width=0.48\textwidth]{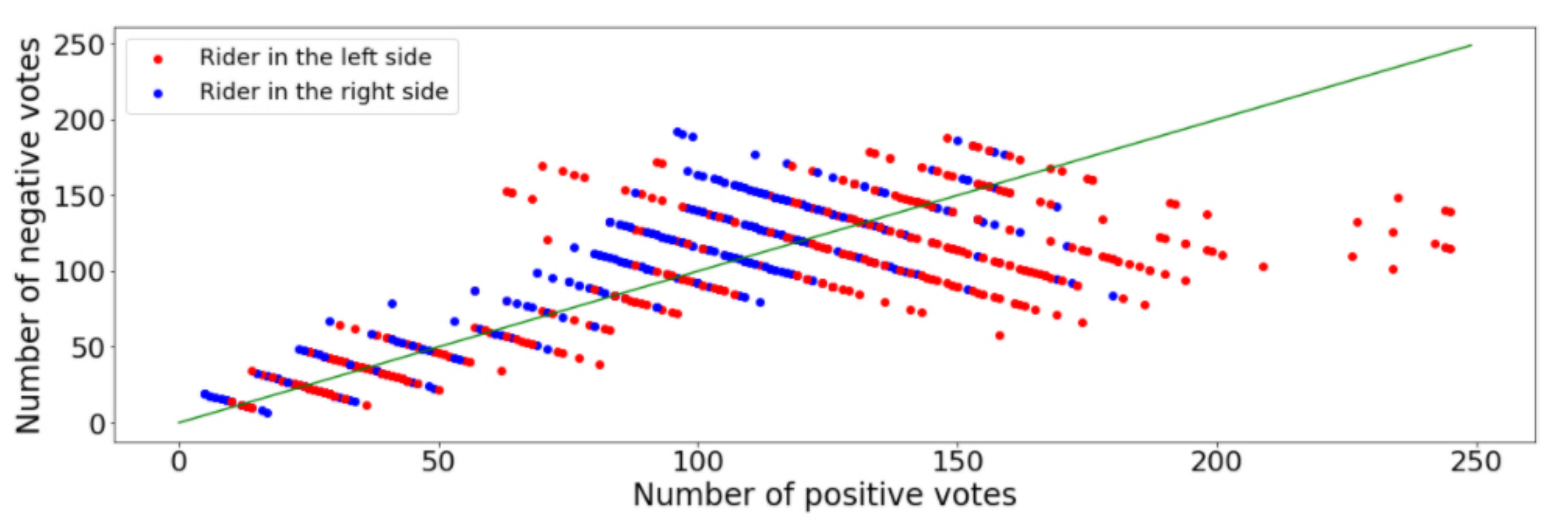}
    \vspace{-0.5em}
    \caption{Positive and negative votes of unwrapped phase difference between antennas when the rider is in the both sides.}
    \vspace{-1em}
    \label{fig:phase-votes}
\end{figure}

\begin{figure}[t]
    \centering
    \includegraphics[width=0.48\textwidth]{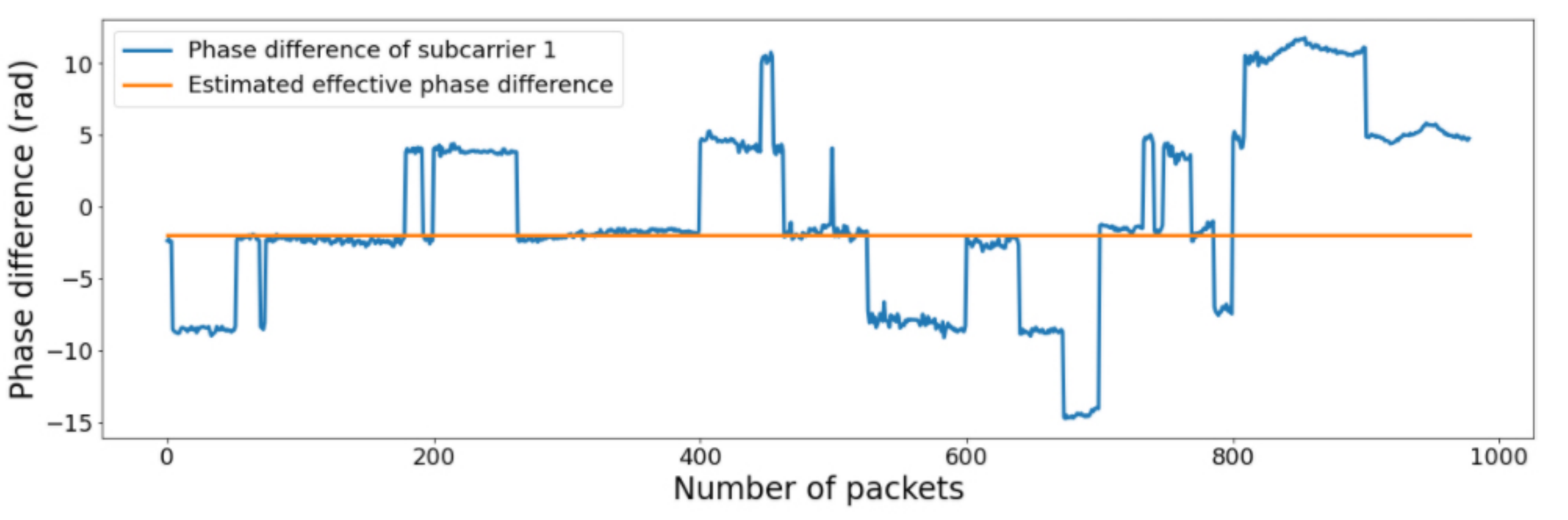}
    \vspace{-0.5em}
    \caption{Estimating effective phase difference between antennas ($C$ - $A$) when the rider is at the right side in LoS.}
    \vspace{-1.5em}
    \label{fig:phase-bar}
\end{figure}

\begin{enumerate}[label=(\alph*)]
    \item We average all phase differences of all sub-carriers of all packets within a window. The intuition is that mean of phase difference should be different for different sides.
    \item Similar to (a), but instead of all the sub-carriers, we use just the first sub-carrier.
    \item We divide the window into a few sub-windows. The reason for sub-windowing is to reduce the propagation error of phase unwrapping. Then, we average all phase differences of all sub-carriers in each sub-window. We remove 20\% sub-windows with large variance. Then, we compute a positive or negative vote for each sub-window based on the sign of its phase difference. We count the numbers of positive and negative votes, and use them as features. We plot the number of positive and negative votes for all the 1032 training windows and plot them in Figure \ref{fig:phase-votes}. The green line shows when the positive and negative votes are the same. It shows that when the rider is on the left (right) side, there are more positive (negative) votes (as they should be). We calculate such votes for all the sub-carriers.
    \item  After sub-windowing, we compute an effective phase difference for each subcarrier. The intuition is that phase difference should be stable for each subcarrier because the central frequency is the same. We choose an effective phase difference that covers most phase differences within two radians and has the smallest mean error. An example of such an effective phase difference is shown in Figure \ref{fig:phase-bar} when the rider is on the right side and in LoS condition. It shows that even though the phase difference fluctuates, the effective phase difference is negative (as it should be since the rider is on the right side). We estimate like this for 30 sub-carriers and use all 30 effective phase differences as features to the classifiers.
\end{enumerate}


We feed these features to kNN, DT, and SVM classifiers and show the results of rider side classification in Table \ref{tab:phase_diff}. For kNN, we vary the value of k from 3 to 15 and report the accuracy with the best k. We see the highest accuracy we get from the phase difference based approach is only 56\%.

\begin{table}[t]
\centering
\caption{Results   of   left   vs.   right   classification of baseline methods.}
\label{tab:phase_diff}
\resizebox{0.75\linewidth}{!}{%
\begin{tabular}{@{}lccc@{}}
\toprule
Baseline & KNN (\%) & DT(\%) & SVM(\%)  \\ \midrule
1(a)    & 50.2 (k=5)    & 46.0   & 48.4                        \\ \midrule
1(b)    & 54.4 (k=11)    & 50.2  & 44.2                       \\ \midrule
1(c)    & 52.6 (k=3)     & 52.6   & 51.6                      \\ \midrule
1(d)    & 52.6 (k=3)    & 56.0   & 49.5     
                  \\ \midrule
                  \midrule
2    & 85.6 (k=10)     & 76.1   & 85.6   
                  \\ \midrule
                  \midrule
3(a)    & 84.2 (k=5)     & 81.1   & 84.6 
                  \\ \midrule
3(b)    & 88.1 (k=3)     & 83.9   &  87.7 
                  \\ \midrule
3(c)    & 83.5 (k=7)     & 81.4   & 85.3  
                  \\ \midrule
3(d)    & 87.3 (k=8)     & 82.5   & 89.5  \\ 
\bottomrule
\end{tabular}%
}
\end{table}

\textbf{Baseline 2: RSS difference based approach:} When we collect data, we also collect RSS (Received Signal Strength) values from each antenna. We feed the average RSS difference of antennas ($C$ - $A$) of each window to different classifiers, including KNN, Decision tree, and SVM, to classify the rider side. The results are shown in Table \ref{tab:phase_diff}. The results show that the highest accuracy is 85.6\% that came from both K-NN and SVM. It provides higher accuracy than the CSI phase difference based approach. 

\textbf{Baseline 3: CSI amplitude difference based approach:}

Since the CSI amplitude difference changes over time, we consider different ways to compute features to capture amplitude difference of antennas ($C$ - $A$):

\begin{enumerate}[label=(\alph*)]
    \item We average all CSI amplitude differences of all sub-carriers of all packets within a window.
    \item Similar to (a), but we use only the first sub-carrier.
    \item Similar to (a), but we also add average RSS difference.
    \item Similar to (b), but we also add average RSS difference.
\end{enumerate}

We feed the features to kNN, DT, and SVM classifiers. The results are shown in Table \ref{tab:phase_diff}. Its shows the highest accuracy is $89.5\%$, when combining the average CSI amplitude difference and average RSS difference.

We also implement our LSTM based network and change network parameters, including the size of hidden dimensions and number of layers to see how that affects performance. The results are shown in Table~\ref{tab:lstm}. It shows that when we use our variance based subcarrier selection, the accuracy is higher than when all sub-carriers are used, or only the first subcarrier is used.
We see that we get 95.44\% accuracy when we combine variance based subcarrier selection, power delay profile, and multipath profile. This highest accuracy came from when we select 14 subcarriers with VbSS, obtain 3 PDP features and 1 multipath profile feature. If we feed the exact same features to kNN, DT, and SVM, we get 68.4\%, 69.5\% , 84.2\% accuracy, respectively. Hence, our LSTM based architecture increases accuracy by 11.24\% from exactly the same input. 

\begin{table*}[t]
\centering
\caption{Results of left vs. right classification when  LSTM  with different features are used.}
\label{tab:lstm}
\resizebox{0.87\linewidth}{!}{%
\begin{tabular}{@{}lccccc@{}}
\toprule
Description                                                                          & Input Dim & Hidden Dim & Number of layers & Optimizer & Accuracy \\ \midrule
Variance-based Subcarrier Selection (VbSS)& 12        & 256        & 3                & RMSProp   & 93.33\%  \\ \midrule
Variance-based Subcarrier Selection (VbSS) & 12        & 256        & 3                & Adam      & 91.57\%  \\ \midrule
Variance-based Subcarrier Selection (VbSS) & 12        & 128        & 3                & RMSProp   & 92.28\%  \\ \midrule
Variance-based Subcarrier Selection (VbSS) & 12        & 256        & 4                & RMSProp   & 89.82\%  \\ \midrule
Variance-based Subcarrier Selection (VbSS) & 12        & 128        & 4                & RMSProp   & 92.63\%  \\ \midrule
First Subcarrier only                                                                & 1         & 256        & 3                & RMSProp   & 89.47\%  \\ \midrule
All sub-carriers                                                                      & 30        & 256        & 3                & RMSProp   & 89.47\%  \\ \midrule
VbSS + Multipath Profile (MP)                                                        & 12+1      & 256        & 3                & RMSProp   & 93.33\%  \\ \midrule
VbSS + Power Delay Profile (PDP)                                                     & 12+5      & 256        & 3                & RMSProp   & 93.33\%  \\ \midrule
VbSS + Power Delay Profile (PDP)                                                     & 12+3      & 256        & 3                & RMSProp   & 93.68\%  \\ \midrule
VbSS + PDP + MP                                                                      & 12+3+1    & 256        & 3                & RMSProp   & 95.08\%  \\ \midrule
VbSS + PDP + MP                                                                      & 14+3+1    & 256        & 3                & RMSProp   & \textbf{95.44\%}  \\ \bottomrule
\end{tabular}%
}
\vspace{-1em}
\end{table*}


\subsection{Sensitivity Analysis}
In this section, we analyze the effect of antenna spacing, subcarrier selection, and window size on \CF performance. 

\textbf{Effect of antenna spacing:}
In our analysis, the default antenna spacing was 5.2 cm, which produced 95.44\% accuracy. Since we collected data with three antennas, we can use antenna $A$ and $B$ to see how the performance looks like when the antenna spacing is 2.6 cm. We keep the best performing network's parameter the same and run the experiment with 2.6 cm spacing and find the accuracy is only 55.79\%. Thus, increasing antenna spacing helps improving accuracy.

\begin{figure}
    \centering
    \includegraphics[width=0.46\textwidth]{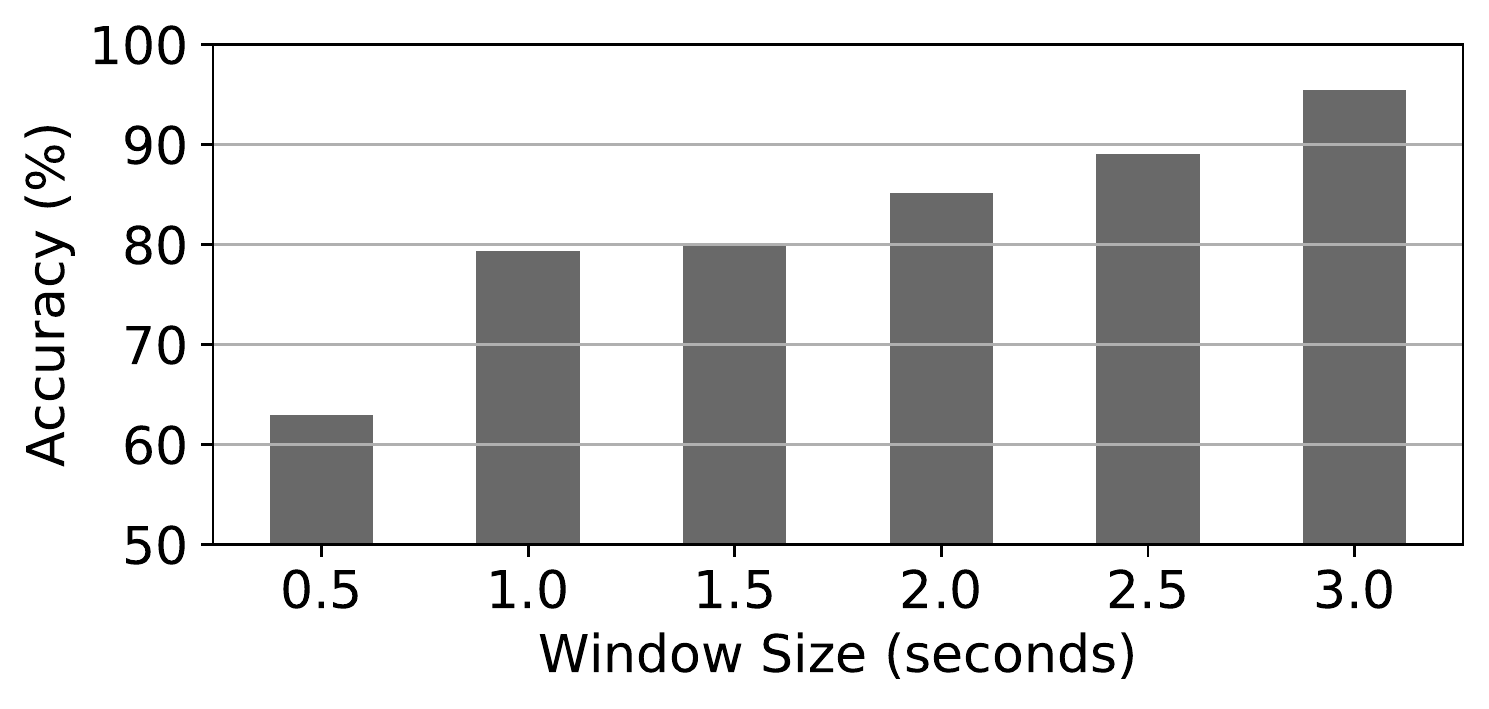}
    \vspace{-0.5em}
    \caption{Effects of window size on accuracy.}
    \vspace{-1.5em}
    \label{fig:window-size}
\end{figure}

\textbf{Effect of window size:}
In our analysis, the default window size is 3 seconds. We keep the best performing network's parameter the same and run the experiment by varying window sizes to 0.5, 1, 1.5, 2, 2.5, and 3 seconds find the accuracy is 62.95\%, 79.36\%, 80\%, 85.17\%, 89.03\%, 95.44\%, respectively as shown in Figure \ref{fig:window-size}. We see that longer windows provide higher accuracy.

\begin{figure}
    \centering
    \includegraphics[width=0.46\textwidth]{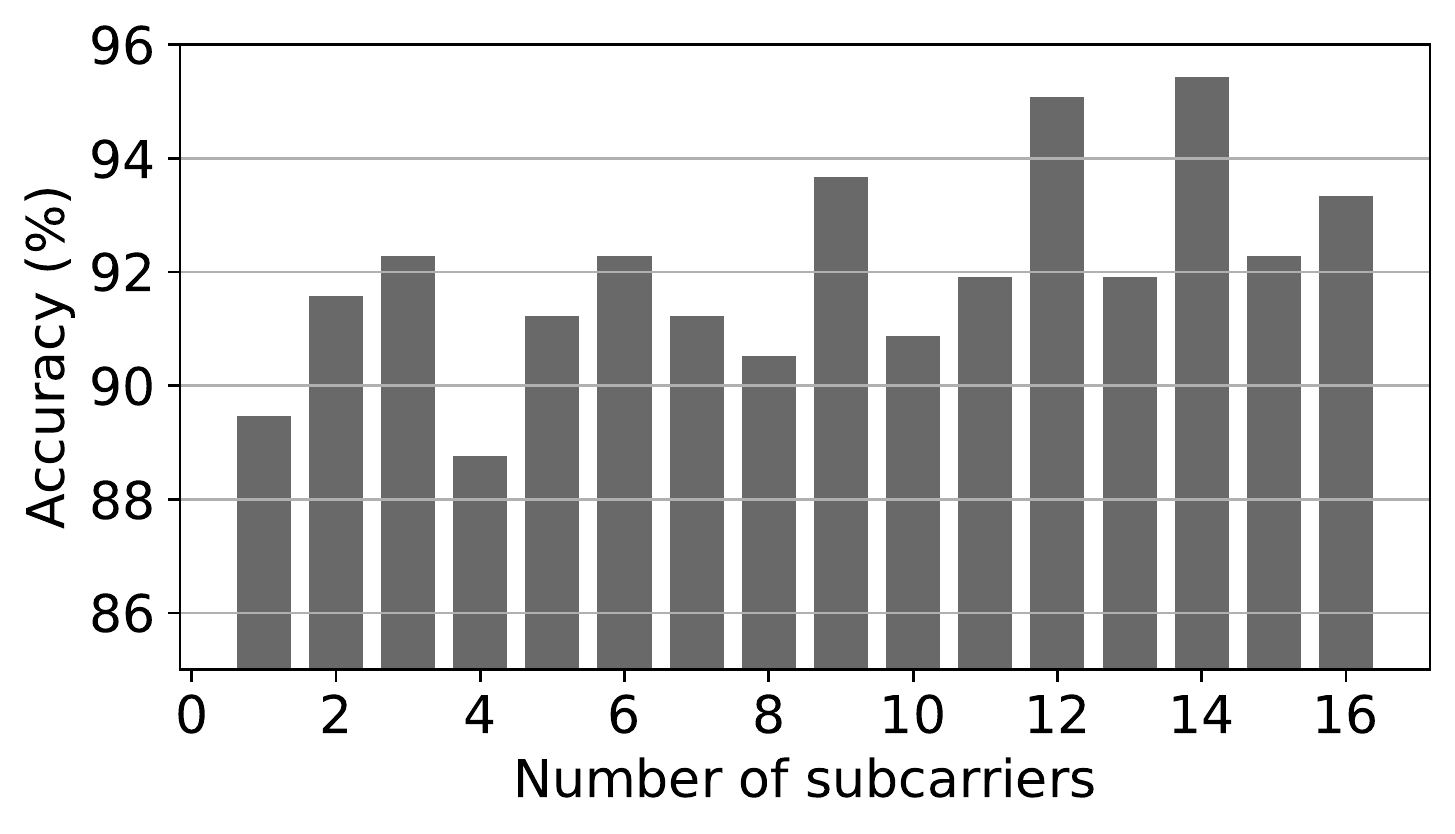}
    \vspace{-0.5em}
    \caption{Effects of number of sub-carriers on accuracy.}
    \vspace{-2em}
    \label{fig:num-sub-carriers}
\end{figure}

\textbf{Effect of number of sub-carriers:} 
We keep the best performing network's parameter the same and run the experiment with changing the number of sub-carriers from 1 to 16 and show the impact of the number of sub-carriers through our VbSS method on accuracy in Figure \ref{fig:num-sub-carriers}. When we choose only one subcarrier, we choose subcarrier 1, as it provides 89.47\% accuracy. We achieve the highest accuracy (95.44\%) when the number of sub-carriers is 14.


%



\subsection{Execution Time}
We train our LSTM using Nvidia GeForce GTX 1080 Ti GPU. It takes about two hours to train the network. However, the inference is rapid. We estimate how long it takes to perform inference in a powerful GPU like NVidia GeForce GTX 1080 Ti as well as an embedded GPU like Nvidia Jetson Nano.  It takes only 101.77 and 850.37 milliseconds to execute the inference process in 1080 Ti and Jetson Nano, respectively. 
Hence, the solution can be run on embedded GPUs in real-time. Also, there are several ways to optimize (e.g., recompiling with TensorRT can significantly reduce inference time on Jetson devices) and prune the model to compress the network, which will reduce inference time \cite{zhu2017prune} \cite{winata2019effectiveness}. We leave this to future work.



\subsection{Range Analysis}
\begin{figure}
    \centering
    \includegraphics[width=0.48\textwidth]{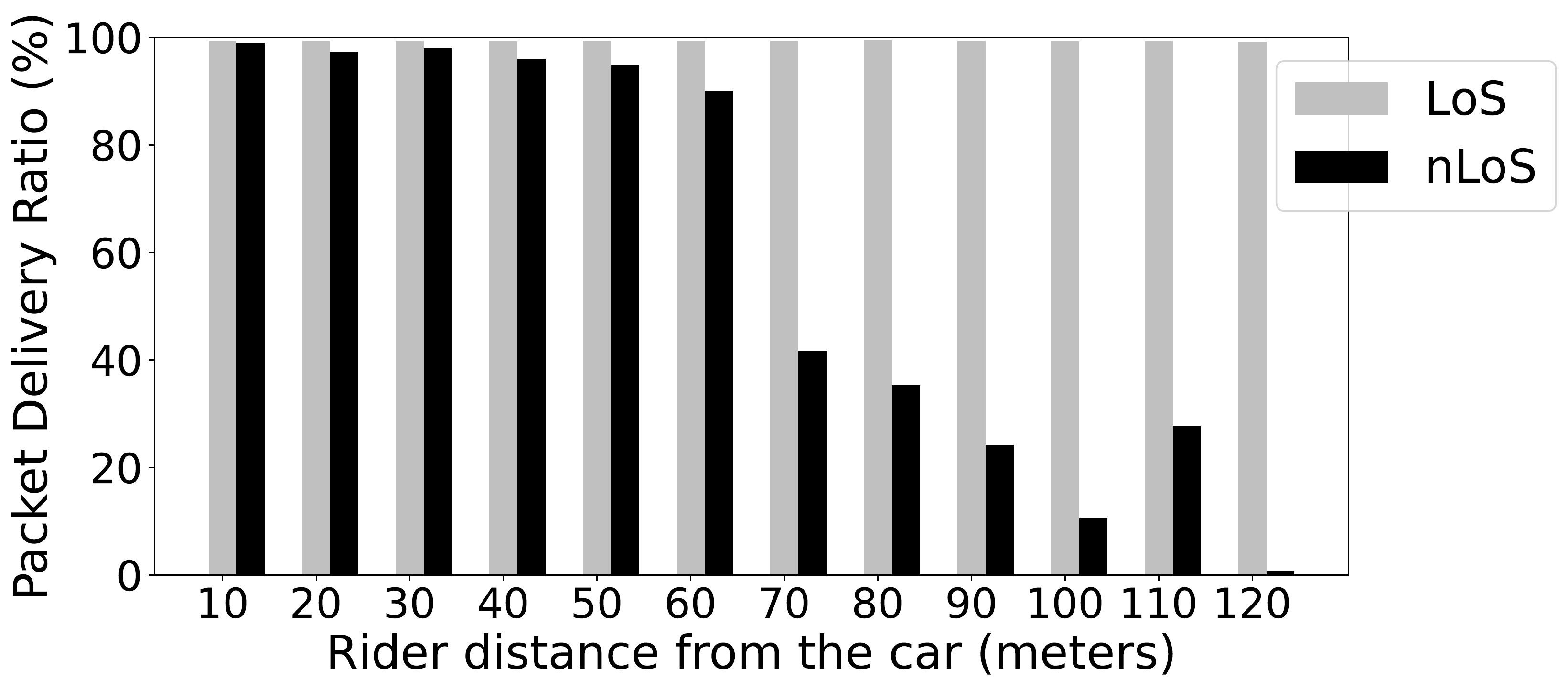}
    \vspace{-0.5em}
    \caption{ Packet delivery ratio at different distances from the car.}
    \vspace{-2em}
    \label{fig:range}
\end{figure}

In this section, we estimate how far \CF can operate in both LoS and nLoS conditions. We collect additional data for this evaluation. We have a person standing at different distances ranging from 10 meters to 120 meters in front of the car in both LoS and nLoS conditions. We transmit 10,003 packets from each location. To create a nLoS condition, we have a person standing between the phone and Wi-Fi receiving unit placed in the car dashboard. The Packet Delivery Ratio (PDR) at different distances from the car is shown in Figure \ref{fig:range}. We see that even at a range of 120 meters, the PDR is 99.30\% in LoS condition. However, in nLoS situation, the PDR drops sharply to 41.65\% in 70 meters. At 120 meters, the PDR is only 0.75\%. So, we see that in LoS conditions, \CF will operate beyond 120 meters range.

%% file: tex/8.discussion.tex
\section{Discussion}

\subsection{Generalizability}
Although the data was collected from one large parking lot, we put an effort to introduce variation in the rides by asking the volunteers to stand differently to block the signal, move while blocking the signal, drive at different speeds, and vary the speed in different rides. As a result, there is a significant variation in the dataset, and we expect the model to generalize to some extent. One particular reason we were not able to collect data from a busy street is that the Wi-Fi of the laptop needed to stay connected to the phone for data collection with CSI Tool \cite{Halperin_csitool}, which is very difficult to obtain in busy streets as the car can easily go out of the Wi-Fi range. Currently, we are switching to Nexmon framework \cite{nexmon:project} for collecting CSI data, where the phone will inject packets at a particular Wi-Fi channel. This will allow us to perform a large-scale data collection from busy streets for testing the generalizability of the solution. We leave it to future work.



\vspace{-0.25em}
\subsection{Limitations}
\vspace{-0.25em}
One potential limitation of our approach is that if someone else books the ride on behalf of the rider and the rider has a different phone, then the proposed solution may not work. If the rider does not co-operate (e.g., by disabling Wi-Fi), then it will not work. Also, it will need support from the rider-hailing service providers (e.g., Uber, Lyft) to enable this service into their apps. However, they can incorporate such a service using their apps and dashboard products. Also, our approach assumes that the rider is waiting for the car on a side of a street while the car is moving towards or away from the rider. If the rider books the trip from inside, e.g.,  from a restaurant, and waits inside for the car to arrive, then we will not be able to leverage the motion related features as effectively, and it may not be able to assist the driver with the rider side. However, when the rider comes out of the restaurant, 
we can ask him to walk along his side of the street to help us capture some features to determine his side. 
Due to the time sensitive nature of the application, assuming it takes maximum 3.85 seconds (0.85 seconds for execution and maximum 3 seconds of window, although 50\% time we only need 1.5 seconds to collect enough packets and perform classification) to run \CF on a Jetson Nano and Wi-Fi range of 70 meters in nLoS, \CF can only determine the rider side when the rider is in front of the car if the vehicle approaches towards the rider at 40.67 miles per hour or less. However, it can work up to 81.34 miles per hour vehicular speed if we are not required to determine rider side while the rider is in front of the car.

\subsection{Lessons Learned and Future Work}
Our exploratory study shows that the phase difference between antennas in an automotive environment is very noisy and does not provide as high accuracy as amplitude difference-based methods for determining rider side. We also find that increasing the antenna distance even by 2.6 cm helps improving accuracy for the amplitude-based approach significantly. The maximum length of an object that can be placed in a dashboard is 5.5 inches by the laws of California, USA. We plan to increase antenna spacing and see if we can achieve higher accuracy at smaller window sizes within this constraint. We leave this to future work. Also, we estimate the rider side independently from each window. In the future, we can take into account all the packets of the past windows from the same ride, assuming that the rider did not cross the street while the car is approaching and have a bigger window for rider side determination. Also, we plan to collect data from busy streets and evaluate the performance in challenging scenarios.

%% file: tex/9.related.tex
\section{Related Work}

\subsection{Human Detection and Identification}

There has been a significant amount of work in detecting the presence of humans and identifying them. 
For the ride-hailing application, the vehicle needs to detect the presence of the rider and identify him/her before localizing him/her. 
Humans can be detected using cameras~\cite{zhu2006fast, wu2011real}, depth sensors~\cite{mithun2018odds, munir2017real, munir2017fork, flores2019dataset, francis2019occutherm, munir2019dataset, liu2017cod, liu2017long}, IR-array sensors~\cite{mohammadmoradi2017measuring},  mmWave radars~\cite{cui2021high, zhao2019mid}, Wi-Fi~\cite{fang2020eyefi, fang2020fusing, fang2020person, chen2022rfcam}, and using other sensors for various purposes~\cite{huang2019magtrack, preum2021review, prabhakara2022exploring}. 

Vision-based systems such as~\cite{wang2018learning, tao2017deep, wang2013insight} can be used to identify the person but requires prior knowledge such as the facial features. These systems are privacy invasive (e.g., facial recognition systems are being banned in multiple cities for privacy concerns) and generally do not work if the subject is at a distance and can be occluded. Systems such as ID-Match~\cite{li2016id} uses RFID tags and a 3D depth camera, FORK \cite{munir2017real} uses a depth sensor, and \cite{llorca2017recognizing} uses RFID and BLE to identify individuals. These works require additional sensors or devices, which are not practical in the ride-hailing scenario, difficult to scale, and potentially costly. Similar to EyeFi~\cite{fang2020eyefi} and RFCam~\cite{chen2022rfcam}, we utilize the phone as an identifier as it is already being utilized in the ride-hailing application.

\subsection{Localization}

Numerous works have focused on localization using Wi-Fi. ~\cite{kotaru2015spotfi} exploits Wi-Fi subcarrier CSI values to create virtual antennas to obtain higher resolution Angle of Arrival (AoA) estimation, then using multiple Wi-Fi receivers to triangulate the transmitter for localization. The usage of multiple Wi-Fi receivers is expensive, and AoA estimation is unstable in our automotive environment. ~\cite{xie2018precise} proposes a higher resolution of power delay profile by utilizing CSI splicing. This method requires a special Wi-Fi configuration, which is impractical for our application. However, it does inspire us to use power delay profile as one of our features. Other works such as  ~\cite{vasisht2016decimeter, wang2011robust, wang2016lifs, soltanaghaei2018multipath} have examined using time-of-flight, frequency, and multipath to estimate AoA and triangulate the locations of the transmitter. These methods are more suited for indoor settings and more prone to environmental changes. 

Using simple features such as RSSI has been explored in previous works to perform vehicle localization~\cite{dinh2017indoor}. Similarly, ~\cite{nguyen2018WiFi} uses a fingerprinting method to localize vehicles in a car park. While fingerprinting is simple and easy to perform, they are prone to environmental changes and lacks generalization ability as it requires prior knowledge for each place. ~\cite{zhang2018WiFi} locates a bus by scanning Wi-Fi APs surrounding the bus route and predicts the time for the bus's arrival. However, it still lacks the generalization ability and can not provide fine-grained location information regarding the rider side.

One work that is closely related to our work is~\cite{ibrahim2020wi}, which uses Wi-Fi Fine Time Measurement (FTM) to measure the distance (through ToF) and achieve high precision localization. However, issues such as the time needed to perform such measurement (which can take several seconds), the requirement of the phone to connect, and the need to place antennas on both sides of the vehicle's roof prevent it from being conveniently deployed. 
Similarly, \cite{pizarro2021accurate} utilizes both CSI and FTM to achieve higher accuracy with added AoA (Angle of Arrival) and AoD (Angle of Departure) measurements. However, this method still depends on FTM  that most of the smartphones do not support.

\subsection{Wi-Fi Sensing}

While some work in Wi-Fi sensing does not directly apply to localization problems, they can provide meaningful features that may be utilized. For example, ~\cite{wu2016widir} use Wi-Fi CSI phase change dynamics with Fresnel zone model to determine the human subject's walking direction. The phase change dynamics contain environmental changes around the transmitter and receiver. In~\cite{jiang2020towards}, the author demonstrated how the Wi-Fi signal contains detailed information that can be used to reconstruct human pose, and a neural network can be used to extract deeper features. In these works, the transmitters and receivers are stationary. However, in our case, the receiver is moving.
